\documentclass[twocolumn,secnumarabic,groupedaddress,superscriptaddress, amssymb, nobibnotes, aps, prd]{revtex4-2}
\usepackage{graphicx}
\pdfoutput=1

\usepackage{color}
\begin{document}


\title{Insulator-to-metal Mott transition facilitated by lattice deformation in monolayer $\alpha$-RuCl$_3$ on graphite}


\author{Xiaohu Zheng}
\email[Correspondence to:] {xhzheng@baqis.ac.cn}
\affiliation{Beijing Academy of Quantum Information Sciences, Beijing 100193, China.}
\author{Ogasawara Takuma}
\affiliation{Beijing Academy of Quantum Information Sciences, Beijing 100193, China.}
\author{Huaxue Zhou}
\affiliation{Beijing Academy of Quantum Information Sciences, Beijing 100193, China.}
\author{Chongli Yang}
\affiliation{Beijing Academy of Quantum Information Sciences, Beijing 100193, China.}
\author{Xin Han}
\affiliation{Beijing National Laboratory for Condensed Matter Physics and Institute of Physics, Chinese Academy of Sciences, Beijing 100190, People's Republic of China.}
\author{Gang Wang}
\affiliation{Department of Microelectronic Science and Engineering, School of Physical Science and Technology, Ningbo University, Ningbo 315211, P. R. China.}
\author{Junhai Ren}
\affiliation{Beijing Academy of Quantum Information Sciences, Beijing 100193, China.}
\author{Youguo Shi}
\affiliation{Beijing National Laboratory for Condensed Matter Physics and Institute of Physics, Chinese Academy of Sciences, Beijing 100190, People's Republic of China.}
\author{Katsumi Tanigaki}
\affiliation{Beijing Academy of Quantum Information Sciences, Beijing 100193, China.}
\author{Rui-Rui Du}
\affiliation{International Center for Quantum Materials, School of Physics, Peking University, Beijing 100871, China.}
\affiliation{CAS Center for Excellence in Topological Quantum Computation, University of Chinese Academy of Sciences, Beijing 100190, China.}
	

\date{\today}

\begin{abstract}
Creating heterostructures with graphene/graphite is a practical method for charge-doping $\alpha$-RuCl$_3$, but not sufficient to cause the insulator-to-metal transition. In this study, detailed scanning tunneling microscopy/spectroscopy measurements on $\alpha$-RuCl$_3$ with various lattice deformations reveal that both in-plane and out-of-plane lattice distortions may collapse the Mott-gap in the case of monolayer $\alpha$-RuCl$_3$ in proximity to graphite, but have little impact on its bulk form alone. In the Mott-Hubbard framework, the transition is attributed to the lattice distortion-facilitated substantial modulation of the electron correlation parameter. Observation of the orbital textures on a highly compressed monolayer $\alpha$-RuCl$_3$ flake on graphite provides valuable evidence that electrons are efficiently transferred from the heterointerface into Cl3$p$ orbitals under the lattice distortion. It is believed that the splitting of Ru $t_{2g}$ bands within the trigonal distortion of Ru-Cl-Ru octahedra bonds generated the electrons transfer pathways. The increase of the Cl3$p$ states enhance the hopping integral in the Mott-Hubbard bands, resulting in the Mott-transition. These findings suggest a new route for implementing the insulator-to-metal transition upon doping in $\alpha$-RuCl$_3$ by deforming the lattice in addition to the formation of heterostructure.
\end{abstract}


\maketitle

\section{INTRODUCTION}
Mott-transition in materials with strong electron correlations is commonly referred to as a discontinuous phase transition (first-order), during which both spin and charge undergo substantial qualitative changes \cite{imada_metal-insulator_1998, sordi_mott_2011, chatzieleftheriou_mott_2023}. It ensures many intriguing physics such as high-temperature superconductivity (HTSC), strange metal phase, charge density wave (CDW), and other symmetry broken states \cite{yee_phase_2015,hanaguri_checkerboard_2004,da_silva_neto_ubiquitous_2014,sakurai_imaging_2011}. Cuprates that exhibit HTSC and charge-order in the vicinity of Mott-transition are an established instance \cite{lee_doping_2006}. Under sufficiently strong spin-orbital-coupling (SOC), Mott insulators can exhibit quantum spin liquid (QSL) phase with topological order and fractional spinon excitation \cite{pesin_mott_2010,broholm_quantum_2020,savary_quantum_2016,zhou_quantum_2017}, providing a conductive environment for investigating a continuous Mott-transition (second-order phase transition) with more intriguing physics \cite{pustogow_quantum_2018,furukawa_quasi-continuous_2018}. By virtue of spin-dependent interactions among spin-half moments \cite{kitaev_anyons_2006,gohlke_quantum_2018,takagi_concept_2019}, the Kitaev honeycomb model admitting an exact QSL state and non-Abelian Majorana spinons stands out as a remarkable example.  Numerous theoretical studies \cite{zhang_phase_2021,you_doping_2012,halasz_doping_2014} have predicted that the Mott-transition in such a Kitaev QSL system could result in a rich phase diagram containing p-wave topological superconducting phase. Charge-doping is regarded as an efficient method for approaching such a Mott-transition. Hitherto, a carrier doped phase of the proximate Kitaev QSL material $\alpha$-RuCl$_3$ has been studied by multiple research groups utilizing various strategies \cite{jo_enhancement_2021, zhou_angle-resolved_2016,zhou_evidence_2019,rizzo_charge-transfer_2020}. Comprising heterostructures of graphene/graphite appear to be the most advantageous among the options reported to date, duo to the ability to prevent the introduction of crystal defects. Experimental evidences have provided empirical support for the notion that electrons, possessing a magnitude of 10$^{13}$ cm$^{-2}$, can be transferred from graphene to $\alpha$-RuCl$_3$, resulting in an equal amount of holes doping in graphene \cite{zhou_evidence_2019,biswas_electronic_2019,mashhadi_spin-split_2019,gerber_ab_2020,souza_magnetic_2022}. However, the insulator-to-metal transition accompanied by the emergence of a well-defined Fermi surface (FS) has not been observed in $\alpha$-RuCl$_3$ yet, impeding further exploration of the potential superconductivity in this system.\par
In our recent study \cite{zheng_tunneling_2023}, we reported that the in-gap states of $\alpha$-RuCl$_3$ in proximity to graphite sensitively experienced a notable augmentation upon the introduction of lattice distortion, which aligns with the theoretical anticipation that straining the lattice may be a powerful potential method for approaching to the Mott-transition in $\alpha$-RuCl$_3$ \cite{iyikanat_tuning_2018,kaib_magnetoelastic_2021,biswas_electronic_2019}. Here, we use scanning tunneling microscopy/spectroscopy (STM/STS) to further explore the electronic properties of $\alpha$-RuCl$_3$ with lattice deformations and strain at liquid nitrogen temperature (77 K). The experimental findings offer a significant clarification that the imposition of strain, regardless of whether it is compressive or extensive, in-plane or out-of-plane, on the lattice of monolayer (ML)-$\alpha$-RuCl$_3$ transferred onto a graphite substrate, may result in the insulator-to-metal Mott transition, while it gives negligible impact on the bulk form of $\alpha$-RuCl$_3$. Upon subjecting the ML-$\alpha$-RuCl$_3$ lattice to a uniform compression of 18\%, we observe atomic-resolved orbital textures, which provide the evidence that newly arising of low-energy electronic states in the Cl3$p$ orbitals are responsible for the Mott-transition. The proposed explanation posits that the lattice distortion has caused the splitting of the $t_{2g}$ band, which provides the pathways for the electrons transferring from the heterointerface to the upper Cl3$p$ orbitals. The new arising states effectively modulate the electron-correlations in Mott-Hubbard bands in $\alpha$-RuCl$_3$. These findings provide a prospective route for the manifestation of FS, thereby affording an opportunity to explore the phenomenon of topological superconductivity in the vicinity of a Mott-transition within a Kitaev QSL.
\section{RESULTS AND DISCUSSION}
\subsection{STM/STS on bulk $\alpha$-RuCl$_3$ with film corrugations}
$\alpha$-RuCl$_3$ is an insulating 4\textit{d} transition-metal halide with honeycomb layers composed of nearly ideal edge-sharing RuCl$_6$ octahedra \cite{plumb_ensuremathalphaensuremath-mathrmrucl_3_2014}, which can be exfoliated into a two-dimensional ML with a thickness of 600 pm \cite{lu_electronic_2023}, as depicted schematically in Fig. 1a. As illustrated in the upper panel of Fig. 1b, the crystal field of RuCl$_6$ octahedra splits the five-fold degenerate \textit{d} levels of 4\textit{d}$^5$ Ru$^{3+}$ into the doublet \textit{e}$_{g}$ and the triplet \textit{t}$_{2g}$ with a charge gap exceeding 2 eV. The \textit{t}$_{2g}$ manifold composed of the \textit{d}$_{xy}$, \textit{d}$_{yz}$ and \textit{d}$_{xz}$ orbitals of Ru$^{3+}$ hybridizes with the 3\textit{p} (\textit{p}$_x$, \textit{p}$_y$, \textit{p}$_z$) orbitals of Cl$^{-}$ \cite{koitzsch_j_eff_2016,iyikanat_tuning_2018}. Subsequently, the SOC (energy: $\lambda=\delta$$_{soc}$) induces an additional splitting within the \textit{t}$_{2g}$ manifold, resulting in a ${J}_{\mathrm{eff}}$=1/2 doublet and a ${J}_{\mathrm{eff}}$=3/2 quartet, characterized by an energy of $\lambda=3\delta$$_{soc}$/2. Consequently, in the hole picture, one hole is accommodated in the lower ${J}_{\mathrm{eff}}$=1/2 Kramer's doublet \cite{takagi_concept_2019,jackeli_mott_2009}, as schematically shown in Fig. 1b. Within the Mott-Hubbard framework, the ${J}_{\mathrm{eff}}$=1/2 manifold undergoes a splitting due to the Coulomb interaction \textit{U}, resulting in the formation of upper Hubbard band (UHB) and lower Hubbard band (LHB). It demonstrates that the Mott-gap can exceed 2 eV, when the value of \textit{U} is set to be 4.5 eV during DFT calculations, as shown in the bottom panel of Fig. 1b, which aligns favorably with the Mott-gap measured via $dI/dV$ spectroscopy and photoemission spectroscopy on bulk $\alpha$-RuCl$_3$ \cite{zheng_tunneling_2023,sinn_electronic_2016}. The results imply that $\alpha$-RuCl$_3$ is a highly correlated electron system.\par
\begin{figure}[htp]
	\centering
	\includegraphics[width=\columnwidth]{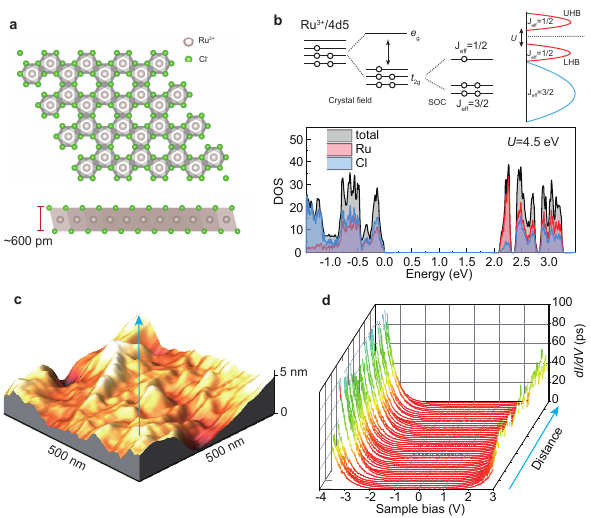}\\
	\linespread{1}
	\caption[Figure 1]{(a) schematically depicts the in-plane and out-plane views of crystal structure of a monolayer $\alpha$-RuCl$_3$, demonstrating that an individual honeycomb layer is formed by edge-sharing RuCl$_6$ octahedra; (b) Upper panel: the level scheme of 4$d^5$ electronic configuration of the ruthenium valence Ru$^{3+}$, where the crystal field and SOC effect together result in a ${J}_{\mathrm{eff}}$=1/2 Mott-Hubbard insulator; Lower panel: DFT calculation of the partial density of states (PDOS) of the $\alpha$-RuCl$_3$ with setting $U$=4.5 eV. The obtained Mott-gap is $\sim$2 eV;(c) STM 3D morphology taken on a thick $\alpha$-RuCl$_3$ with clear corrugation of the film being observed, bias voltage: $V_b$=2.5 V, setpoint current: $I_s$=500 pA; (d) $dI/dV$ spectra collected along the green line in (c) shows the Mott-Hubbard bands is robust. The spectra are measured with parameters: $V_b$=2.5 V and $I_s$=500 pA; lock-in frequency of 707 Hz, and amplitude modulation of 5 mV.}
	\label{fig:figure-1}
\end{figure}
Several academic studies \cite{bu_possible_2019,dymkowski_strain-induced_2014} have demonstrated that the imposition of strain or pressure upon the crystal of a Mott insulator possesses the potential to cause an insulator-to-metal Mott-transition. Our prior research \cite{zheng_tunneling_2023} has revealed that a minor distortion of lattice in ML-$\alpha$-RuCl$_3$ can substantially modify the electronic characteristic by causing the in-gap states within the Mott-gap. A question arises whether the applied strain or the lattice distortion possesses the capacity to trigger a Mott-transition within the bulk form of $\alpha$-RuCl$_3$. In this study, we exfoliated the $\alpha$-RuCl$_3$ thin flakes using the Scotch tapes from a bulk single crystal synthesized from commercially available RuCl$_3$ powder by means of a vacuum sublimation. After repeatedly reducing the thickness of the film, we finally exfoliated a target $\alpha$-RuCl$_3$ thin flakes using a thermal release tap in order to transfer to a fresh surface of highly oriented pyrolytic graphite  (HOPG) substrate. After heating the sample to 120 $^o$C for 20 seconds in an argon atmosphere in the glovebox, the $\alpha$-RuCl$_3$ thin films were released on the graphite surface. Lattice deformation and strain of the $\alpha$-RuCl$_3$ film are inevitably introduced during the transfer process. Then, the sample was directly transferred into an STM chamber and annealed at 280 $^o$C in an ultra-high vacuum chamber (1E-10 Torr) for at least 2 hours for degassing and improving the contacting quality prior to STM/STS measurements. We measured the thicknesses of $\alpha$-RuCl$_3$ flakes by a scanning topographic image at the step-edge and measured its height-profile to the graphite surface. We started by examining the condition of a thick $\alpha$-RuCl$_3$ flake in presence of the pronounced film corrugations, as demonstrated by the STM morphology in Fig. 1c. Through the acquisition of \textit{dI/dV} spectra along the designated path indicated by the blue arrow line in Fig. 1c, it ascertained that the large Mott-gap, measured approximately to be 2 eV, remains robust over the corrugations. In contrast to the case of ML-$\alpha$-RuCl$_3$ on graphite in our previous work where the in-gap states arise in the Mott-gap\cite{zheng_tunneling_2023}, the corrugations did not give rise to any in-gap states, as shown in Fig. 1d. This serves to validate the notion that a considerable crystal deformation is incapable of collapsing the strong Mott-Hubbard framework in bulk form $\alpha$-RuCl$_3$. The viewpoint is also supported by previous calculations that the Mott-Hubbard band of $\alpha$-RuCl$_3$ is hardly changed under a biaxial in-plane tensile strain of 8\% \cite{vatansever_strain_2019}.\par
\subsection {Mott-gap collapsed by the modest strain in ML-$\alpha$-RuCl$_3$ in proximity to graphite}
Subsequently, our focus was redirected towards the characteristics exhibited by the ML-$\alpha$-RuCl$_3$ following its transfer onto the graphite. As depicted in the STM image in Fig. 2a, a selected region of ML-$\alpha$-RuCl$_3$ exhibits substantial film corrugations. The STM morphology was subjected to the fast Fourier transform (FFT), and the observable shifts of the Bragg peaks associated with the Ru and Cl atoms indicate that the out-of-plane corrugation has caused the in-plane distortion of the crystal lattices, as shown inset of Fig. 2a. Previous studies have revealed that the lattice morphology under STM within the sample bias (\textit{V}$_{b}$) range of -3.0 to 1.5 V is dominated by Ru \textit{t}$_{2g}$ and Cl \textit{p}$_\pi$ (\textit{p}$_{z}$) orbitals, and the resulting hybridization of these orbitals gives rise to the characteristic corner-shared Kagome pattern in STM images \cite{wang_identification_2022}, which has been demonstrated in our recent experiment on bulk $\alpha$-RuCl$_3$ \cite{zheng_tunneling_2023}. Nevertheless, upon observation on ML-$\alpha$-RuCl$_3$, it was noted that the FFT-filtered image (as shown in Fig. 2b) did not exhibit the anticipated Kagome-like patterns. This suggests that forming heterostructure has exerted a significant influence on the orbital configurations in ML-$\alpha$-RuCl$_3$, a finding that seems to contradict the observations made on the epitaxial ML-$\alpha$-RuCl$_3$ on graphite \cite{wang_identification_2022}.\par
\begin{figure}[htp]
	\centering
	\includegraphics[width=\columnwidth]{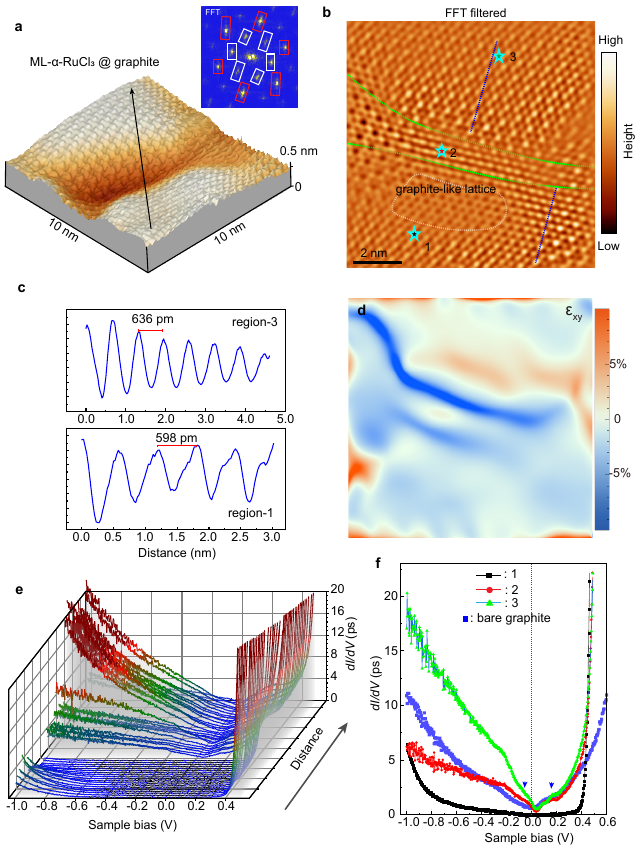}\\
		\linespread{1}
	\caption[Figure 2]{(a) STM atomic-resolved morphology of ML-$\alpha$-RuCl$_3$, where corrugation in the film is clearly visible; inset show a FFT of the image, where the Bragg peaks corresponding to the Ru (white squares) and Cl (red squares) atom sites, and the strain-induced displacements of the Bragg peaks can be observed; (b) FFT filtered STM image shows that the surface is divided into three regions (denoted as region-1, -2 and -3) by the corrugation, which present different lattice strains; (c) height profiles along the dashes in (b) display that the lattice in region-1 has negligible distortion with a lattice constant of a= 6 \AA, region-2 compressed, and region-3 significantly expanded with the lattice constant of a= 6.36 \AA; (d) Strain field map obtained from geometric phase analysis represented by the $\epsilon_{xy}$ components of the strain tensors; (e) \textit{dI/dV} spectra collected from the spanned surface show the evolution of band structures at 77 K, where the Mott-Hubbard gap is preserved in region-1, and the Mott-gaps are collapsed in regions -2 and -3 with lattice distortions ($V_{b}$=500 mV and $I_s$=500 pA); (f) the representative \textit{dI/dV} spectra acquired at 77 K from the three different regions, respectively. The blue curve underneath these spectra is \textit{dI/dV} spectrum from bare graphite.}
	\label{fig:figure-2}
\end{figure}
The film corrugation partitions the surface into three separate regions: region-1, characterized by a nearly strain-free lattice; region-2, where the lattice experiences compression; and region-3, where the lattice undergoes expansion, as seen in Fig. 2c. In region-1, an intriguing observation is the existence of an area exhibiting graphene-like lattices. This observation aligns with our previous research \cite{zheng_tunneling_2023}, which suggests the possibility of hetero-interfacial hybridization between $\alpha$-RuCl$_3$ and graphite. The lattice constants for regions-1 and -3 were determined by extracting the height profile along the dashed lines, yielding the values of approximately 6 \AA\, for region-1 and 6.4 \AA\, for region-3. The strain distribution across the surface has been conducted utilizing the geometric phase analysis method \cite{hytch_quantitative_1998}, as shown in Fig. 2d. Subsequently, \textit{dI/dV} spectra were obtained along the trajectory illustrated in Fig. 2a, encompassing the three aforementioned regions. The results display that the large Mott-gap maintains itself in region-1, whereas the Mott-gap collapses within the presence of a finite local density of states (LDOS) at the Fermi level in regions-2 and -3. Irrespective of whether the material experiences compressive or tensile strain, as depicted by the strain field maps in Fig. 2d, the findings suggest that the lattice distortions induce the emergence of FS in $\alpha$-RuCl$_3$. Given the negligible impact of comparable lattice distortions on the bulk film, as depicted in Figs. 1c and d, it is highly indicative that the lattice distortions have effectively facilitated the emergence of in-gap states in ML-$\alpha$-RuCl$_3$ within the heterostructure.The observed phenomenon is consistance with the theoretical calculation that applying pressure on the $\alpha$-RuCl$_3$/graphene heterostructure enables the interfacial charge transfer in $\alpha$-RuCl$_3$ \cite{gerber_ab_2020}. Through the quantitative analysis with regard to the spectra observed within the three defined regions in Fig. 2f, it was discovered that, despite the compressed region (region-2) displaying a reduced occupied LDOS in comparison to the region subjected to tensile strain (region-3), the line-shape of the spectra within both regions exhibited a remarkable similarity, rendering them nearly indiscernible. All these observed metallic \textit{dI/dV} spectra exhibit a notable dip near the Fermi level, being reminiscent of the pseudogap observed in the charge-doped cuprate.\par
\begin{figure}[htp]
	\centering
	\includegraphics[width=\columnwidth]{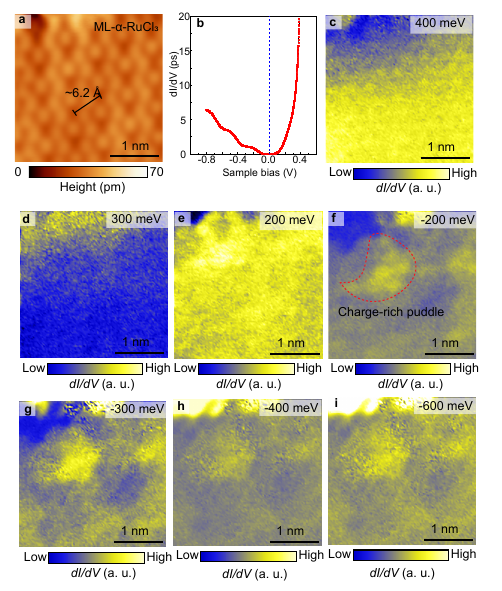}\\
	\linespread{1}
	\caption[Figure 3]{(a) atomic-resolved STM image collected on the  ML-$\alpha$-RuCl$_3$ with expanded lattice; (b) averaged \textit{dI/dV} spectrum acquired on the surface of (a), $V_b$=400 mV, $I_s$=0.5 nA; (c to i) energy-dependent \textit{dI/dV} maps with variation of the bias energies from 400 meV to -600 meV ($V_b$=400 mV, $I_s$=1 nA; T=77 K; Lock-in: 707 Hz, 8 mV).}
	\label{fig:figure-3}
\end{figure}
Then, we measured energy-resolved \textit{dI/dV} maps on a ML-$\alpha$-RuCl$_3$ with its lattice under modest tension ($\sim$3.5\%), as depicted in Fig. 3a. Fig. 3b displays the average spectral curve, which has a nearly closed energy gap. Although the tensile strain has indeed resulted in an enhancement of the distinct states within the Mott-gap, our observations in \textit{dI/dV} maps (figs. 3c to i) have not revealed any discernible contrasts in orbital texture. This outcome aligns with our previous findings on strain-free ML-$\alpha$-RuCl$_3$\cite{zheng_tunneling_2023}. It is implied that although the electrons present in $\alpha$-RuCl$_3$ possess equal magnitudes to the doped holes in graphite, approximately 10$^{13}$ cm$^{-2}$, these electrons have not occupied any orbitals within the energy range being examined by STS under the condition of modest lattice distortion. Consequently, it implies that the electrons in the in-gap states are itinerant or have been strongly dispersed during the tunneling process. Charge puddles can be seen in \textit{dI/dV} maps that are acquired under negative energies, as shown in Figs. 3e-i. These charge puddles are thought to arise from the inhomogeneous distribution of the electrons, which aligns with the recent theoretical prediction that the spatial distribution of the doping electrons exhibits inhomogeneity in $\alpha$-RuCl$_3$ \cite{souza_magnetic_2022}. Similar charge puddles are also frequently observed in charge-doped cuprates before the emergence of charge-ordered phase \cite{cai_visualizing_2016,battisti_universality_2017,da_silva_neto_doping-dependent_2016,frano_charge_2020}.\par
\begin{figure*}[htp]
	\centering
	\includegraphics[width=2\columnwidth]{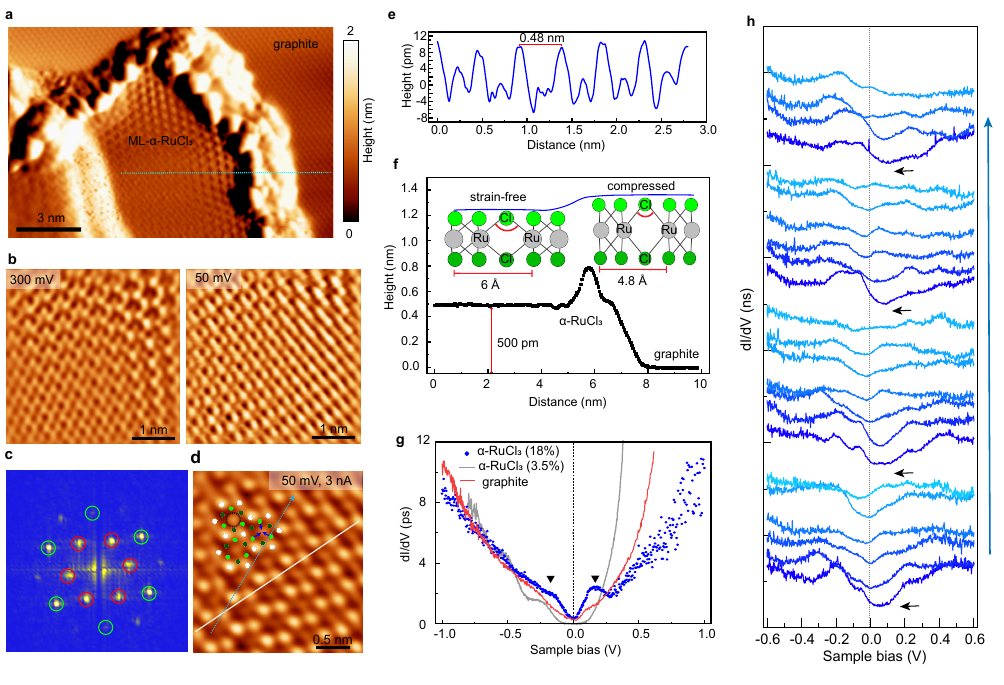}\\
	\linespread{1}
	\caption[Figure 4]{(a) STM morphology of a ML-$\alpha$-RuCl$_3$ flake on graphite ($V_b$=500 mV; $I_s$=200 pA); (b) atomic-resolved STM images taken at different sample bias ( $V_b$=300 mV and 50 mV, $I_s$=1 nA) on $\alpha$-RuCl$_3$; (c) FFT image obtained from (b), where the Bragg peaks associated with Ru and Cl atom sites maintain the hexagonal symmetry that same to the strain-free $\alpha$-RuCl$_3$ film; (d) high-resolution STM morphology taken with a very small tip-surface distance setting by $V_b$=50 mV, $I_s$=3 nA; (e) line-profile along the white dashed line in (d) demonstrates the lattice is significantly compressed with a lattice constant a= 4.8 \AA, being shrunk by 18\%; (f) height profile along the green line in (a) presenting the thickness of the film; (g) comparison of averaged $dI/dV$ spectra collected on the ML-$\alpha$-RuCl$_3$ with large lattice compression ($\sim$18\%), modest tension ($\sim$3.5\%) and bare graphite at 77 K, respectively ($V_b$=500 mV and $I_s$=200 pA); (h) \textit{dI/dV} spectra along the green dash in (d) spanning over several honeycomb lattices which reveal the atomic-resolved evolution of the LDOS on the surface ($V_b$=100 mV, $I_s$=1 nA; T=77 K; Lock-in:707 Hz, 8 mV).}
	\label{fig:figure-4}
\end{figure*}
\subsection {Energy dependent orbital textures in the extremely strained ML-$\alpha$-RuCl$_3$}
To further elucidate the intricate interplay among lattice deformation, charge transfer, and the insulator-to-metal transition within the ML-$\alpha$-RuCl$_3$ in proximity to graphite, we examined another extraordinary scenario wherein substantial biaxial strain (the lattice is shrunk over 18\%) is unintentionally imposed upon a piece of ML-$\alpha$-RuCl$_3$ flake, as illustrated in Fig. 4a. Atomic-resolved STM images acquired at various sample biases, as depicted in Fig. 4b, show a lattice morphology that exhibits similar features when the lattice is subjected to modest tensile strain, as illustrated in Fig. 2c. The Kagome-like pattern as well as the trimer pattern expected in pristine $\alpha$-RuCl$_3$ were also not detected at the low energy scale, which is a supplementary evidence notifying that the orbital configurations of ML-$\alpha$-RuCl$_3$ are indeed modified upon lattice deformation and being in proximity to graphite. Moreover, the examined flake also failed to exhibit a graphite-like lattice under a bias as small as 50 mV (Fig. 4b), which contradicts the observations depicted in strain-free scenarios \cite{zheng_tunneling_2023}. Therefore, it indicates that the electron tunneling procedure through the heterostructure has been more seriously modified.\par
The FFT image of the STM morphology displays the Bragg peaks corresponding to the Ru and Cl atoms, which maintains the hexagonal lattice symmetry under the strain, as depicted in Fig. 4c. By decreasing the distance between STM tip and the surface, hexagonal Ru atoms of the $\alpha$-RuCl$_3$ lattice becomes clear, as shown in Fig. 4d. Following the STM profile, Fig. 4e showcases that the lattice constant is 4.8 \AA\, in this strained flake, which elucidates that the lattice is subjected to a substantial compression, exceeding -18\%. The application of such a large in-plane compressive strain amplifies the out-of-plane buckling. Thickness of the present flake (500 pm), as displayed in Fig. 4f, is greater than that of the strain-free flake (350 pm) obtained in our previous investigation \cite{zheng_tunneling_2023}. It can be imagined that the significant compression in the lattice results from the thermal treatment employed during the fabrication of the heterostructure. As seen in the averaged \textit{dI/dV} spectrum of the compressed lattice, as shown in Fig. 4g, the Mott-gap is thoroughly collapsed. It results in an emergence of a similar pseudogap-like feature (within the range of $\pm$20 meV) near the Fermi level, akin to the ones observed in Fig. 2f. These empirical observations suggest that the application of strains may exert a systematic influence on the Mott-transition. Spatially resolved \textit{dI/dV} spectra along the blue arrow line in Fig. 4d exhibit repetitive modulation of the line-shapes that aligns with the lattice's symmetry, as depicted in Fig. 4h. This observation implies the existence of distinct orbital textures that align with the lattice geometry, potentially unveiling insights into the underlying mechanism of the strain facilitated Mott-transition.\par
\begin{figure*}[htp]
	\centering
	\includegraphics[width=1.5\columnwidth]{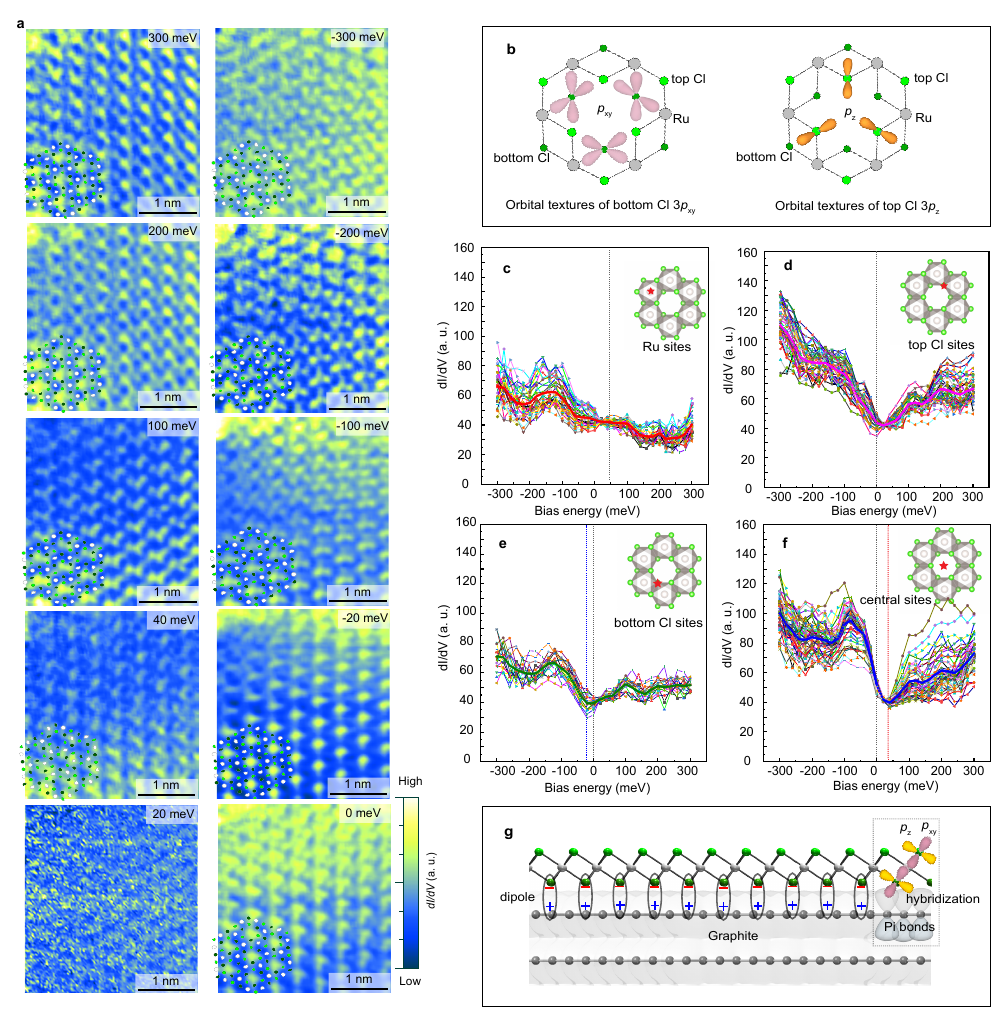}\\
	\linespread{1}
	\caption[Figure 5]{(a) energy-dependent \textit{dI/dV} maps taken at the same region as shown in Fig. 4d depict the orbital textures that evolve as the changing of the energy levels. The lattice structural models are schematically put on the surfaces ($V_b$=50 mV, $I_s$=0.6 nA; T=77 K; Lock-in: 707 Hz, 8 mV); (b) schematic Cl3$p$ orbital configurations of $p_{xy}$ and $p_z$ (possible orbital textures under $dI/dV$ mapping) in the ML-$\alpha$-RuCl$_3$; (c) to (f) show the representative averaged \textit{dI/dV} spectra collected at different sites (c: Ru-sites, d: top-Cl sites, e: bottom-Cl sites and f: center of honeycombs, as pointed out by the red stars on the lattice models) of the lattice from $dI/dV$ maps in (a); (g) schematic interface dipoles formed between bottom-located Cl and graphite surface, which is induced by the hybridization (charge transfer) between Cl3$p$ orbitals and the $\pi$ orbitals of graphite surface.}
	\label{fig:figure-5}
\end{figure*}
Therefore, we conducted an analysis of energy-resolved \textit{dI/dV} mappings within the defined region depicted in Fig. 4d, encompassing an energy range spanning from 300 mV to -300 mV, as visually represented in Fig. 5a. In contrast to the depicted scenario in Fig. 3, wherein the orbital texture appears indistinct, the observed region showcases remarkably discernible energy-dependent and atomically-resolved orbital textures. This observation indicates that these low-energy surface electron states reside in specified orbitals within the system. The obtained \textit{dI/dV} maps within the energy range reveal a remarkably diminished LDOS on Ru sites, in constrast to the pronounced concentration of high intensity in the proximity of Cl atoms. The manifestation of a weak orbital texture arising from the hybridization of Ru and Cl orbitals (\textit{t}$_{2g}$-\textit{p}$_{z}$ hybridization) is confined to an energy magnitude of 40 meV. The fact that the LDOS at Cl sites exhibited a network-like pattern at the energy levels exceeding 200 meV is concurrent with a depletion of Ru sites, indicating that the pattern is predominantly composed by the Cl3$p$ orbitals. Two distinct patterns of orbital textures are discernible when examining the occupied states. In the vicinity of Fermi level, precisely at 0 and -20 meV of the \textit{dI/dV} maps, the LDOSs exhibit pronounced intensity patterns situated in the central region of the hexagonal lattice. In the absence of solid atomic entities occupying the central positions within $\alpha$-RuCl$_3$ lattice, we hypothesize that the central LDOSs are predominantly generated from the $\pi$ bonds present on the graphite surface. Below -20 meV, the configuration of a three-pointed star within the hexagonal lattice can be observed. It can be unequivocally attributed to the 3\textit{p}$_{z}$ lobes of the topmost Cl atoms, since they are primarily detected by STM tip, and contribute the three-pointed star patterns to the orbital textures, as illustrated in Fig. 5b.\par
The aforementioned observations demonstrate that the Cl3\textit{p} orbitals are the primary source of the textures at a low energies, which is paradoxical given that the low-energy electronic configuration of the $\alpha$-RuCl$_3$ could be dominated by the degenerate $e_g$ and $t_{2g}$ bands from the Ru4\textit{d} orbitals, whereas the Cl3\textit{p} orbital electrons predominantly occupy the valence band below -2 eV \cite{lu_electronic_2023,wang_identification_2022,sinn_electronic_2016}. In fact, the current observations are in agreement with the theoretical computations, which indicate that the electrons transferred from the heterointerface have a greater affinity for bonding to the Cl sites \cite{souza_magnetic_2022}. It suggests that the observed electron states in the orbital texture are the result of charge transfer in the heterostructure. In addition, the strain in ML-$\alpha$-RuCl$_3$ has altered the electron distribution from the heterointerface to the Cl orbitals as a result of the lattice distortion.\par
The spatially-averaged LDOS was then collected at distinct locations, specifically the Ru sites, the top and bottom Cl sites, and the sites in the center of the hexagons, as depicted in Figs. 5c to f. The LDOS curves observed at the Ru sites exhibit a nearly flat profile with a relatively reduced intensity (Fig. 5c). This feature implies that the Coulomb reputation still wishes to preserve the Mott-gap in $t_{2g}$ band, and also explains the depletion morphology observed on the Ru sites in \textit{dI/dV} maps in Fig. 5a. The spectral data collected from the top and bottom Cl sites as well as the center of the hexagons all exhibit a pseudogap-like feature in the spectral curves with a minimum value near the Fermi level (as depicted in Figs. 5d and f). It is noteworthy that the curves originating from the bottom Cl sites present a comparatively lower intensity due to the increased tunneling distance and a negative shift of its minimum point. The spectra measured at the centers of the hexagons exhibit an asymmetric line shape, with the occupied states being significantly higher than the empty ones. In addition, as depicted in Fig. 5f, a positive shift of the spectral minimum is observed. As was previously mentioned, the central points have potential to provide information regarding graphite. Comparing the \textit{dI/dV} spectra to those of bare graphite, the positive shift of the minimum can be interpreted as evidence of hole-doping, which is consistent with previous findings that the graphite in the heterostructure is hole doped \cite{rizzo_charge-transfer_2020,rossi_direct_2023,zhou_evidence_2019,biswas_electronic_2019}. The negative shift of the spectral minimum at bottom Cl sites is consistent with the band calculations \cite{souza_magnetic_2022} that the low-energy bands of $\alpha$-RuCl$_3$ shift downward in the heterostructure. As depicted schematically in Fig. 5g, an opposite shift of the spatial spectra suggests the presence of an interfacial dipole layer between $\alpha$-RuCl$_3$ and graphene \cite{rossi_direct_2023}. The hybridization between Cl3$p$ and $\pi$ orbitals of graphite adequately accounts for the formation of interface dipole.\par
This extreme case with significant lattice compression allows us to visualize the orbital structures in ML-$\alpha$-RuCl$_3$ in close proximity to graphite and provides information regarding the interfacial dipoles. However, it remains unclear how the lattice distortion could cause the Mott-Hubbard band to collapse. It is essential to note, even though a large amount of electrons is transferred from graphite to strain-free ML-$\alpha$-RuCl$_3$ in the heterostructure, the substantial Mott-gap ($\sim$2 eV)\cite{zheng_tunneling_2023} may effectively impede the movements of electrons from heterointerface into the Mott-Hubbard band. Therefore, forming the dipole layer is resonable, as depicted in Fig. 5f. Such a propose can accurately interpret the findings in our previous studies that the in-gap states were observable as the probing tip approaching to the surface \cite{zheng_tunneling_2023}, and that the orbital textures on the top surface were indistinct (Fig. 3). However, by referring to the Jahn-Teller theorem and the discussion in a recent study concerning on epitaxially grown $\alpha$-RuCl$_3$ on graphite \cite{wang_identification_2022}, any lattice distortion to prolong (shorten) the Ru-Ru bonds within widening (reducing) the Ru-Cl-Ru angles would cause the trigonal distortion of the Ru-Cl octahedra, as depicted schematically in Fig. 4f (inset). Such distortion will cause the splitting of \textit{t}$_{2g}$ manifold into \textit{a}$_{1g}$ singlet and \textit{e}$_g^{'}$ doublet, while preserving the hybridization with the $p_z$ orbitals of both the top and bottom Cl atoms. Despite the fact that $\alpha$-RuCl$_3$ bulk and ML remain in the Mott insulating state under \textit{t}$_{2g}$ splitting \cite{Liu_PhysRevB.107.165134,vatansever_strain_2019}, the distribution of doped electrons $\alpha$-RuCl$_3$ may be ultimately altered. Therefore, it is proposed with sufficient adequacy that the electrons residing at the bottom Cl layer may exhibit a higher propensity to transfer into the top Cl orbitals via the band splitting and the hybridization of $p_z^{bottom}$-$a_{1g}$-$p_z^{top}$ and $p_z^{bottom}$-$e_g^{'}$-$p_z^{top}$. The redistribution of electrons concurrently decreases the parameter $U/t$ ($t$ is the hopping integral), or breaks the half-filled $t_{2g}$ band in the Mott-Hubbard framework, which collapses the Mott-gap as shown in the \textit{dI/dV} spectra in the strained ML-$\alpha$-RuCl$_3$.  
\section{CONCLUSION}
In conclusion, we reported the observation of lattice-distortion facilitated Mott-gap collapse in ML-$\alpha$-RuCl$_3$ in the heterostructure involving graphite. We clarified that the identical film deformation failed to change the Mott-gap in bulk $\alpha$-RuCl$_3$. Important information was provided by the energy-resolved orbital textures by conducting STM/STS measurements on an ML-$\alpha$-RuCl$_3$ flake experiencing an extremely large compressed lattice strain. It was proven that a majority of charges for collapsing the Mott-gap reside in the Cl3$p$ orbitals. In order to comprehend the physical mechanism, a model was proposed that the strain causes the splitting of $t_{2g}$ band into $a_{1g}$ and $e_g^{'}$ orbitals, and the orbital hybridization of ($a_{1g}$ and $e_g^{'}$) with the Cl\textit{p}$_{z}$ greatly modify the distribution of electrons those are transferred from graphite. Part of electrons previously accumulated at the herterointerface (bottom Cl layer) are transferred to the upper Cl layers, via the $p_z^{bottom}$-$a_{1g}$-$p_z^{top}$ and $p_z^{bottom}$-$e_g^{'}$-$p_z^{top}$ pathways under lattice deformation, resulting in the Mott transition in the contexts of the doped Mott-Hubbard bands. The results will provide avenues for investigating the topological superconductivity in the vicinity of a Mott-transition within a Kitaev QSL candidate. 

\begin{acknowledgments}
This work was supported by National Basic Research and Development plan of China (Grants No. 2019YFA0308400), the National Natural Science Foundation of China (Grants No. U2032204, 12174027, 62174093), and the Strategic Priority Research Program of the Chinese Academy of Sciences (Grants No. XDB28000000 and XDB33030000).
\end{acknowledgments}
\bibliography{Maintext-revised.bib}

\begin{thebibliography}{47}%
\makeatletter
\providecommand \@ifxundefined [1]{%
 \@ifx{#1\undefined}
}%
\providecommand \@ifnum [1]{%
 \ifnum #1\expandafter \@firstoftwo
 \else \expandafter \@secondoftwo
 \fi
}%
\providecommand \@ifx [1]{%
 \ifx #1\expandafter \@firstoftwo
 \else \expandafter \@secondoftwo
 \fi
}%
\providecommand \natexlab [1]{#1}%
\providecommand \enquote  [1]{``#1''}%
\providecommand \bibnamefont  [1]{#1}%
\providecommand \bibfnamefont [1]{#1}%
\providecommand \citenamefont [1]{#1}%
\providecommand \href@noop [0]{\@secondoftwo}%
\providecommand \href [0]{\begingroup \@sanitize@url \@href}%
\providecommand \@href[1]{\@@startlink{#1}\@@href}%
\providecommand \@@href[1]{\endgroup#1\@@endlink}%
\providecommand \@sanitize@url [0]{\catcode `\\12\catcode `\$12\catcode
  `\&12\catcode `\#12\catcode `\^12\catcode `\_12\catcode `\%12\relax}%
\providecommand \@@startlink[1]{}%
\providecommand \@@endlink[0]{}%
\providecommand \url  [0]{\begingroup\@sanitize@url \@url }%
\providecommand \@url [1]{\endgroup\@href {#1}{\urlprefix }}%
\providecommand \urlprefix  [0]{URL }%
\providecommand \Eprint [0]{\href }%
\providecommand \doibase [0]{https://doi.org/}%
\providecommand \selectlanguage [0]{\@gobble}%
\providecommand \bibinfo  [0]{\@secondoftwo}%
\providecommand \bibfield  [0]{\@secondoftwo}%
\providecommand \translation [1]{[#1]}%
\providecommand \BibitemOpen [0]{}%
\providecommand \bibitemStop [0]{}%
\providecommand \bibitemNoStop [0]{.\EOS\space}%
\providecommand \EOS [0]{\spacefactor3000\relax}%
\providecommand \BibitemShut  [1]{\csname bibitem#1\endcsname}%
\let\auto@bib@innerbib\@empty
\bibitem [{\citenamefont {Imada}\ \emph {et~al.}(1998)\citenamefont {Imada},
  \citenamefont {Fujimori},\ and\ \citenamefont
  {Tokura}}]{imada_metal-insulator_1998}%
  \BibitemOpen
  \bibfield  {author} {\bibinfo {author} {\bibfnamefont {M.}~\bibnamefont
  {Imada}}, \bibinfo {author} {\bibfnamefont {A.}~\bibnamefont {Fujimori}},\
  and\ \bibinfo {author} {\bibfnamefont {Y.}~\bibnamefont {Tokura}},\
  }\href@noop {} {\bibfield  {journal} {\bibinfo  {journal} {Review of Modern
  Physics}\ }\textbf {\bibinfo {volume} {70}},\ \bibinfo {pages} {1039}
  (\bibinfo {year} {1998})}\BibitemShut {NoStop}%
\bibitem [{\citenamefont {Sordi}\ \emph {et~al.}(2011)\citenamefont {Sordi},
  \citenamefont {Haule},\ and\ \citenamefont {Tremblay}}]{sordi_mott_2011}%
  \BibitemOpen
  \bibfield  {author} {\bibinfo {author} {\bibfnamefont {G.}~\bibnamefont
  {Sordi}}, \bibinfo {author} {\bibfnamefont {K.}~\bibnamefont {Haule}},\ and\
  \bibinfo {author} {\bibfnamefont {A.-M.~S.}\ \bibnamefont {Tremblay}},\
  }\href {https://doi.org/https://doi.org/10.1103/PhysRevB.84.075161}
  {\bibfield  {journal} {\bibinfo  {journal} {Physical Review B}\ }\textbf
  {\bibinfo {volume} {84}},\ \bibinfo {pages} {075161} (\bibinfo {year}
  {2011})}\BibitemShut {NoStop}%
\bibitem [{\citenamefont {Chatzieleftheriou}\ \emph {et~al.}(2023)\citenamefont
  {Chatzieleftheriou}, \citenamefont {Kowalski}, \citenamefont {Berović},
  \citenamefont {Amaricci}, \citenamefont {Capone}, \citenamefont {De~Leo},
  \citenamefont {Sangiovanni},\ and\ \citenamefont {de’
  Medici}}]{chatzieleftheriou_mott_2023}%
  \BibitemOpen
  \bibfield  {author} {\bibinfo {author} {\bibfnamefont {M.}~\bibnamefont
  {Chatzieleftheriou}}, \bibinfo {author} {\bibfnamefont {A.}~\bibnamefont
  {Kowalski}}, \bibinfo {author} {\bibfnamefont {M.}~\bibnamefont {Berović}},
  \bibinfo {author} {\bibfnamefont {A.}~\bibnamefont {Amaricci}}, \bibinfo
  {author} {\bibfnamefont {M.}~\bibnamefont {Capone}}, \bibinfo {author}
  {\bibfnamefont {L.}~\bibnamefont {De~Leo}}, \bibinfo {author} {\bibfnamefont
  {G.}~\bibnamefont {Sangiovanni}},\ and\ \bibinfo {author} {\bibfnamefont
  {L.}~\bibnamefont {de’ Medici}},\ }\href
  {https://doi.org/10.1103/PhysRevLett.130.066401} {\bibfield  {journal}
  {\bibinfo  {journal} {Physical Review Letters}\ }\textbf {\bibinfo {volume}
  {130}},\ \bibinfo {pages} {066401} (\bibinfo {year} {2023})},\ \bibinfo
  {note} {publisher: American Physical Society}\BibitemShut {NoStop}%
\bibitem [{\citenamefont {Yee}\ and\ \citenamefont
  {Balents}(2015)}]{yee_phase_2015}%
  \BibitemOpen
  \bibfield  {author} {\bibinfo {author} {\bibfnamefont {C.-H.}\ \bibnamefont
  {Yee}}\ and\ \bibinfo {author} {\bibfnamefont {L.}~\bibnamefont {Balents}},\
  }\href {https://doi.org/https://doi.org/10.1103/PhysRevX.5.021007} {\bibfield
   {journal} {\bibinfo  {journal} {Physical Review X}\ }\textbf {\bibinfo
  {volume} {5}},\ \bibinfo {pages} {021007} (\bibinfo {year}
  {2015})}\BibitemShut {NoStop}%
\bibitem [{\citenamefont {Hanaguri}\ \emph {et~al.}(2004)\citenamefont
  {Hanaguri}, \citenamefont {Lupien}, \citenamefont {Kohsaka}, \citenamefont
  {Lee}, \citenamefont {Azuma}, \citenamefont {Takano}, \citenamefont
  {Takagi},\ and\ \citenamefont {Davis}}]{hanaguri_checkerboard_2004}%
  \BibitemOpen
  \bibfield  {author} {\bibinfo {author} {\bibfnamefont {T.}~\bibnamefont
  {Hanaguri}}, \bibinfo {author} {\bibfnamefont {C.}~\bibnamefont {Lupien}},
  \bibinfo {author} {\bibfnamefont {Y.}~\bibnamefont {Kohsaka}}, \bibinfo
  {author} {\bibfnamefont {D.-H.}\ \bibnamefont {Lee}}, \bibinfo {author}
  {\bibfnamefont {M.}~\bibnamefont {Azuma}}, \bibinfo {author} {\bibfnamefont
  {M.}~\bibnamefont {Takano}}, \bibinfo {author} {\bibfnamefont
  {H.}~\bibnamefont {Takagi}},\ and\ \bibinfo {author} {\bibfnamefont {J.~C.}\
  \bibnamefont {Davis}},\ }\href {https://doi.org/10.1038/nature02861}
  {\bibfield  {journal} {\bibinfo  {journal} {Nature}\ }\textbf {\bibinfo
  {volume} {430}},\ \bibinfo {pages} {1001} (\bibinfo {year} {2004})},\
  \bibinfo {note} {number: 7003 Publisher: Nature Publishing Group}\BibitemShut
  {NoStop}%
\bibitem [{\citenamefont {da~Silva~Neto}\ \emph {et~al.}(2014)\citenamefont
  {da~Silva~Neto}, \citenamefont {Aynajian}, \citenamefont {Frano},
  \citenamefont {Comin}, \citenamefont {Schierle}, \citenamefont {Weschke},
  \citenamefont {Gyenis}, \citenamefont {Wen}, \citenamefont {Schneeloch},
  \citenamefont {Xu}, \citenamefont {Ono}, \citenamefont {Gu}, \citenamefont
  {Le~Tacon},\ and\ \citenamefont {Yazdani}}]{da_silva_neto_ubiquitous_2014}%
  \BibitemOpen
  \bibfield  {author} {\bibinfo {author} {\bibfnamefont {E.~H.}\ \bibnamefont
  {da~Silva~Neto}}, \bibinfo {author} {\bibfnamefont {P.}~\bibnamefont
  {Aynajian}}, \bibinfo {author} {\bibfnamefont {A.}~\bibnamefont {Frano}},
  \bibinfo {author} {\bibfnamefont {R.}~\bibnamefont {Comin}}, \bibinfo
  {author} {\bibfnamefont {E.}~\bibnamefont {Schierle}}, \bibinfo {author}
  {\bibfnamefont {E.}~\bibnamefont {Weschke}}, \bibinfo {author} {\bibfnamefont
  {A.}~\bibnamefont {Gyenis}}, \bibinfo {author} {\bibfnamefont
  {J.}~\bibnamefont {Wen}}, \bibinfo {author} {\bibfnamefont {J.}~\bibnamefont
  {Schneeloch}}, \bibinfo {author} {\bibfnamefont {Z.}~\bibnamefont {Xu}},
  \bibinfo {author} {\bibfnamefont {S.}~\bibnamefont {Ono}}, \bibinfo {author}
  {\bibfnamefont {G.}~\bibnamefont {Gu}}, \bibinfo {author} {\bibfnamefont
  {M.}~\bibnamefont {Le~Tacon}},\ and\ \bibinfo {author} {\bibfnamefont
  {A.}~\bibnamefont {Yazdani}},\ }\href
  {https://doi.org/10.1126/science.1243479} {\bibfield  {journal} {\bibinfo
  {journal} {Science}\ }\textbf {\bibinfo {volume} {343}},\ \bibinfo {pages}
  {393} (\bibinfo {year} {2014})},\ \bibinfo {note} {publisher: American
  Association for the Advancement of Science}\BibitemShut {NoStop}%
\bibitem [{\citenamefont {Sakurai}\ \emph {et~al.}(2011)\citenamefont
  {Sakurai}, \citenamefont {Itou}, \citenamefont {Barbiellini}, \citenamefont
  {Mijnarends}, \citenamefont {Markiewicz}, \citenamefont {Kaprzyk},
  \citenamefont {Gillet}, \citenamefont {Wakimoto}, \citenamefont {Fujita},
  \citenamefont {Basak}, \citenamefont {Wang}, \citenamefont {Al-Sawai},
  \citenamefont {Lin}, \citenamefont {Bansil},\ and\ \citenamefont
  {Yamada}}]{sakurai_imaging_2011}%
  \BibitemOpen
  \bibfield  {author} {\bibinfo {author} {\bibfnamefont {Y.}~\bibnamefont
  {Sakurai}}, \bibinfo {author} {\bibfnamefont {M.}~\bibnamefont {Itou}},
  \bibinfo {author} {\bibfnamefont {B.}~\bibnamefont {Barbiellini}}, \bibinfo
  {author} {\bibfnamefont {P.~E.}\ \bibnamefont {Mijnarends}}, \bibinfo
  {author} {\bibfnamefont {R.~S.}\ \bibnamefont {Markiewicz}}, \bibinfo
  {author} {\bibfnamefont {S.}~\bibnamefont {Kaprzyk}}, \bibinfo {author}
  {\bibfnamefont {J.-M.}\ \bibnamefont {Gillet}}, \bibinfo {author}
  {\bibfnamefont {S.}~\bibnamefont {Wakimoto}}, \bibinfo {author}
  {\bibfnamefont {M.}~\bibnamefont {Fujita}}, \bibinfo {author} {\bibfnamefont
  {S.}~\bibnamefont {Basak}}, \bibinfo {author} {\bibfnamefont {Y.~J.}\
  \bibnamefont {Wang}}, \bibinfo {author} {\bibfnamefont {W.}~\bibnamefont
  {Al-Sawai}}, \bibinfo {author} {\bibfnamefont {H.}~\bibnamefont {Lin}},
  \bibinfo {author} {\bibfnamefont {A.}~\bibnamefont {Bansil}},\ and\ \bibinfo
  {author} {\bibfnamefont {K.}~\bibnamefont {Yamada}},\ }\href
  {https://doi.org/10.1126/science.1199391} {\bibfield  {journal} {\bibinfo
  {journal} {Science}\ }\textbf {\bibinfo {volume} {332}},\ \bibinfo {pages}
  {698} (\bibinfo {year} {2011})},\ \bibinfo {note} {publisher: American
  Association for the Advancement of Science}\BibitemShut {NoStop}%
\bibitem [{\citenamefont {Lee}\ \emph {et~al.}(2006)\citenamefont {Lee},
  \citenamefont {Nagaosa},\ and\ \citenamefont {Wen}}]{lee_doping_2006}%
  \BibitemOpen
  \bibfield  {author} {\bibinfo {author} {\bibfnamefont {P.~A.}\ \bibnamefont
  {Lee}}, \bibinfo {author} {\bibfnamefont {N.}~\bibnamefont {Nagaosa}},\ and\
  \bibinfo {author} {\bibfnamefont {X.-G.}\ \bibnamefont {Wen}},\ }\href
  {https://doi.org/10.1103/RevModPhys.78.17} {\bibfield  {journal} {\bibinfo
  {journal} {Reviews of Modern Physics}\ }\textbf {\bibinfo {volume} {78}},\
  \bibinfo {pages} {17} (\bibinfo {year} {2006})},\ \bibinfo {note} {publisher:
  American Physical Society}\BibitemShut {NoStop}%
\bibitem [{\citenamefont {Pesin}\ and\ \citenamefont
  {Balents}(2010)}]{pesin_mott_2010}%
  \BibitemOpen
  \bibfield  {author} {\bibinfo {author} {\bibfnamefont {D.}~\bibnamefont
  {Pesin}}\ and\ \bibinfo {author} {\bibfnamefont {L.}~\bibnamefont
  {Balents}},\ }\href {https://doi.org/10.1038/nphys1606} {\bibfield  {journal}
  {\bibinfo  {journal} {Nature Physics}\ }\textbf {\bibinfo {volume} {6}},\
  \bibinfo {pages} {376} (\bibinfo {year} {2010})},\ \bibinfo {note} {number: 5
  Publisher: Nature Publishing Group}\BibitemShut {NoStop}%
\bibitem [{\citenamefont {Broholm}\ \emph {et~al.}(2020)\citenamefont
  {Broholm}, \citenamefont {Cava}, \citenamefont {Kivelson}, \citenamefont
  {Nocera}, \citenamefont {Norman},\ and\ \citenamefont
  {Senthil}}]{broholm_quantum_2020}%
  \BibitemOpen
  \bibfield  {author} {\bibinfo {author} {\bibfnamefont {C.}~\bibnamefont
  {Broholm}}, \bibinfo {author} {\bibfnamefont {R.~J.}\ \bibnamefont {Cava}},
  \bibinfo {author} {\bibfnamefont {S.~A.}\ \bibnamefont {Kivelson}}, \bibinfo
  {author} {\bibfnamefont {D.~G.}\ \bibnamefont {Nocera}}, \bibinfo {author}
  {\bibfnamefont {M.~R.}\ \bibnamefont {Norman}},\ and\ \bibinfo {author}
  {\bibfnamefont {T.}~\bibnamefont {Senthil}},\ }\href
  {https://doi.org/10.1126/science.aay0668} {\bibfield  {journal} {\bibinfo
  {journal} {Science}\ }\textbf {\bibinfo {volume} {367}},\ \bibinfo {pages}
  {eaay0668} (\bibinfo {year} {2020})},\ \bibinfo {note} {publisher: American
  Association for the Advancement of Science Section: Review}\BibitemShut
  {NoStop}%
\bibitem [{\citenamefont {Savary}\ and\ \citenamefont
  {Balents}(2016)}]{savary_quantum_2016}%
  \BibitemOpen
  \bibfield  {author} {\bibinfo {author} {\bibfnamefont {L.}~\bibnamefont
  {Savary}}\ and\ \bibinfo {author} {\bibfnamefont {L.}~\bibnamefont
  {Balents}},\ }\href {https://doi.org/10.1088/0034-4885/80/1/016502}
  {\bibfield  {journal} {\bibinfo  {journal} {Reports on Progress in Physics}\
  }\textbf {\bibinfo {volume} {80}},\ \bibinfo {pages} {016502} (\bibinfo
  {year} {2016})},\ \bibinfo {note} {publisher: IOP Publishing}\BibitemShut
  {NoStop}%
\bibitem [{\citenamefont {Zhou}\ \emph {et~al.}(2017)\citenamefont {Zhou},
  \citenamefont {Kanoda},\ and\ \citenamefont {Ng}}]{zhou_quantum_2017}%
  \BibitemOpen
  \bibfield  {author} {\bibinfo {author} {\bibfnamefont {Y.}~\bibnamefont
  {Zhou}}, \bibinfo {author} {\bibfnamefont {K.}~\bibnamefont {Kanoda}},\ and\
  \bibinfo {author} {\bibfnamefont {T.-K.}\ \bibnamefont {Ng}},\ }\href
  {https://doi.org/10.1103/RevModPhys.89.025003} {\bibfield  {journal}
  {\bibinfo  {journal} {Reviews of Modern Physics}\ }\textbf {\bibinfo {volume}
  {89}},\ \bibinfo {pages} {025003} (\bibinfo {year} {2017})},\ \bibinfo {note}
  {publisher: American Physical Society}\BibitemShut {NoStop}%
\bibitem [{\citenamefont {Pustogow}\ \emph {et~al.}(2018)\citenamefont
  {Pustogow}, \citenamefont {Bories}, \citenamefont {Löhle}, \citenamefont
  {Rösslhuber}, \citenamefont {Zhukova}, \citenamefont {Gorshunov},
  \citenamefont {Tomić}, \citenamefont {Schlueter}, \citenamefont {Hübner},
  \citenamefont {Hiramatsu}, \citenamefont {Yoshida}, \citenamefont {Saito},
  \citenamefont {Kato}, \citenamefont {Lee}, \citenamefont {Dobrosavljević},
  \citenamefont {Fratini},\ and\ \citenamefont
  {Dressel}}]{pustogow_quantum_2018}%
  \BibitemOpen
  \bibfield  {author} {\bibinfo {author} {\bibfnamefont {A.}~\bibnamefont
  {Pustogow}}, \bibinfo {author} {\bibfnamefont {M.}~\bibnamefont {Bories}},
  \bibinfo {author} {\bibfnamefont {A.}~\bibnamefont {Löhle}}, \bibinfo
  {author} {\bibfnamefont {R.}~\bibnamefont {Rösslhuber}}, \bibinfo {author}
  {\bibfnamefont {E.}~\bibnamefont {Zhukova}}, \bibinfo {author} {\bibfnamefont
  {B.}~\bibnamefont {Gorshunov}}, \bibinfo {author} {\bibfnamefont
  {S.}~\bibnamefont {Tomić}}, \bibinfo {author} {\bibfnamefont {J.~A.}\
  \bibnamefont {Schlueter}}, \bibinfo {author} {\bibfnamefont {R.}~\bibnamefont
  {Hübner}}, \bibinfo {author} {\bibfnamefont {T.}~\bibnamefont {Hiramatsu}},
  \bibinfo {author} {\bibfnamefont {Y.}~\bibnamefont {Yoshida}}, \bibinfo
  {author} {\bibfnamefont {G.}~\bibnamefont {Saito}}, \bibinfo {author}
  {\bibfnamefont {R.}~\bibnamefont {Kato}}, \bibinfo {author} {\bibfnamefont
  {T.-H.}\ \bibnamefont {Lee}}, \bibinfo {author} {\bibfnamefont
  {V.}~\bibnamefont {Dobrosavljević}}, \bibinfo {author} {\bibfnamefont
  {S.}~\bibnamefont {Fratini}},\ and\ \bibinfo {author} {\bibfnamefont
  {M.}~\bibnamefont {Dressel}},\ }\href
  {https://doi.org/10.1038/s41563-018-0140-3} {\bibfield  {journal} {\bibinfo
  {journal} {Nature Materials}\ }\textbf {\bibinfo {volume} {17}},\ \bibinfo
  {pages} {773} (\bibinfo {year} {2018})},\ \bibinfo {note} {number: 9
  Publisher: Nature Publishing Group}\BibitemShut {NoStop}%
\bibitem [{\citenamefont {Furukawa}\ \emph {et~al.}(2018)\citenamefont
  {Furukawa}, \citenamefont {Kobashi}, \citenamefont {Kurosaki}, \citenamefont
  {Miyagawa},\ and\ \citenamefont {Kanoda}}]{furukawa_quasi-continuous_2018}%
  \BibitemOpen
  \bibfield  {author} {\bibinfo {author} {\bibfnamefont {T.}~\bibnamefont
  {Furukawa}}, \bibinfo {author} {\bibfnamefont {K.}~\bibnamefont {Kobashi}},
  \bibinfo {author} {\bibfnamefont {Y.}~\bibnamefont {Kurosaki}}, \bibinfo
  {author} {\bibfnamefont {K.}~\bibnamefont {Miyagawa}},\ and\ \bibinfo
  {author} {\bibfnamefont {K.}~\bibnamefont {Kanoda}},\ }\href
  {https://doi.org/10.1038/s41467-017-02679-7} {\bibfield  {journal} {\bibinfo
  {journal} {Nature Communications}\ }\textbf {\bibinfo {volume} {9}},\
  \bibinfo {pages} {307} (\bibinfo {year} {2018})},\ \bibinfo {note} {number: 1
  Publisher: Nature Publishing Group}\BibitemShut {NoStop}%
\bibitem [{\citenamefont {Kitaev}(2006)}]{kitaev_anyons_2006}%
  \BibitemOpen
  \bibfield  {author} {\bibinfo {author} {\bibfnamefont {A.}~\bibnamefont
  {Kitaev}},\ }\href {https://doi.org/10.1016/j.aop.2005.10.005} {\bibfield
  {journal} {\bibinfo  {journal} {Annals of Physics}\ }\textbf {\bibinfo
  {volume} {321}},\ \bibinfo {pages} {2} (\bibinfo {year} {2006})},\ \bibinfo
  {note} {arXiv: cond-mat/0506438}\BibitemShut {NoStop}%
\bibitem [{\citenamefont {Gohlke}\ \emph {et~al.}(2018)\citenamefont {Gohlke},
  \citenamefont {Wachtel}, \citenamefont {Yamaji}, \citenamefont {Pollmann},\
  and\ \citenamefont {Kim}}]{gohlke_quantum_2018}%
  \BibitemOpen
  \bibfield  {author} {\bibinfo {author} {\bibfnamefont {M.}~\bibnamefont
  {Gohlke}}, \bibinfo {author} {\bibfnamefont {G.}~\bibnamefont {Wachtel}},
  \bibinfo {author} {\bibfnamefont {Y.}~\bibnamefont {Yamaji}}, \bibinfo
  {author} {\bibfnamefont {F.}~\bibnamefont {Pollmann}},\ and\ \bibinfo
  {author} {\bibfnamefont {Y.~B.}\ \bibnamefont {Kim}},\ }\href
  {https://doi.org/10.1103/PhysRevB.97.075126} {\bibfield  {journal} {\bibinfo
  {journal} {Physical Review B}\ }\textbf {\bibinfo {volume} {97}},\ \bibinfo
  {pages} {075126} (\bibinfo {year} {2018})},\ \bibinfo {note} {publisher:
  American Physical Society}\BibitemShut {NoStop}%
\bibitem [{\citenamefont {Takagi}\ \emph {et~al.}(2019)\citenamefont {Takagi},
  \citenamefont {Takayama}, \citenamefont {Jackeli}, \citenamefont
  {Khaliullin},\ and\ \citenamefont {Nagler}}]{takagi_concept_2019}%
  \BibitemOpen
  \bibfield  {author} {\bibinfo {author} {\bibfnamefont {H.}~\bibnamefont
  {Takagi}}, \bibinfo {author} {\bibfnamefont {T.}~\bibnamefont {Takayama}},
  \bibinfo {author} {\bibfnamefont {G.}~\bibnamefont {Jackeli}}, \bibinfo
  {author} {\bibfnamefont {G.}~\bibnamefont {Khaliullin}},\ and\ \bibinfo
  {author} {\bibfnamefont {S.~E.}\ \bibnamefont {Nagler}},\ }\href
  {https://doi.org/10.1038/s42254-019-0038-2} {\bibfield  {journal} {\bibinfo
  {journal} {Nature Reviews Physics}\ }\textbf {\bibinfo {volume} {1}},\
  \bibinfo {pages} {264} (\bibinfo {year} {2019})},\ \bibinfo {note} {number: 4
  Publisher: Nature Publishing Group}\BibitemShut {NoStop}%
\bibitem [{\citenamefont {Zhang}\ and\ \citenamefont
  {Liu}(2021)}]{zhang_phase_2021}%
  \BibitemOpen
  \bibfield  {author} {\bibinfo {author} {\bibfnamefont {S.-M.}\ \bibnamefont
  {Zhang}}\ and\ \bibinfo {author} {\bibfnamefont {Z.-X.}\ \bibnamefont
  {Liu}},\ }\href {https://doi.org/10.1103/PhysRevB.104.115108} {\bibfield
  {journal} {\bibinfo  {journal} {Physical Review B}\ }\textbf {\bibinfo
  {volume} {104}},\ \bibinfo {pages} {115108} (\bibinfo {year} {2021})},\
  \bibinfo {note} {publisher: American Physical Society}\BibitemShut {NoStop}%
\bibitem [{\citenamefont {You}\ \emph {et~al.}(2012)\citenamefont {You},
  \citenamefont {Kimchi},\ and\ \citenamefont {Vishwanath}}]{you_doping_2012}%
  \BibitemOpen
  \bibfield  {author} {\bibinfo {author} {\bibfnamefont {Y.-Z.}\ \bibnamefont
  {You}}, \bibinfo {author} {\bibfnamefont {I.}~\bibnamefont {Kimchi}},\ and\
  \bibinfo {author} {\bibfnamefont {A.}~\bibnamefont {Vishwanath}},\ }\href
  {https://doi.org/10.1103/PhysRevB.86.085145} {\bibfield  {journal} {\bibinfo
  {journal} {Physical Review B}\ }\textbf {\bibinfo {volume} {86}},\ \bibinfo
  {pages} {085145} (\bibinfo {year} {2012})},\ \bibinfo {note} {publisher:
  American Physical Society}\BibitemShut {NoStop}%
\bibitem [{\citenamefont {Hal\'asz}\ \emph {et~al.}(2014)\citenamefont
  {Hal\'asz}, \citenamefont {Chalker},\ and\ \citenamefont
  {Moessner}}]{halasz_doping_2014}%
  \BibitemOpen
  \bibfield  {author} {\bibinfo {author} {\bibfnamefont {G.~B.}\ \bibnamefont
  {Hal\'asz}}, \bibinfo {author} {\bibfnamefont {J.~T.}\ \bibnamefont
  {Chalker}},\ and\ \bibinfo {author} {\bibfnamefont {R.}~\bibnamefont
  {Moessner}},\ }\href {https://doi.org/10.1103/PhysRevB.90.035145} {\bibfield
  {journal} {\bibinfo  {journal} {Physical Review B}\ }\textbf {\bibinfo
  {volume} {90}},\ \bibinfo {pages} {035145} (\bibinfo {year} {2014})},\
  \bibinfo {note} {publisher: American Physical Society}\BibitemShut {NoStop}%
\bibitem [{\citenamefont {Jo}\ \emph {et~al.}(2021)\citenamefont {Jo},
  \citenamefont {Heo}, \citenamefont {Lee}, \citenamefont {Choi}, \citenamefont
  {Kim}, \citenamefont {Jeong}, \citenamefont {Jeong}, \citenamefont {Yuk},
  \citenamefont {Eom}, \citenamefont {Jahng}, \citenamefont {Lee},
  \citenamefont {Jung}, \citenamefont {Cho}, \citenamefont {Kim}, \citenamefont
  {Cho}, \citenamefont {Kang},\ and\ \citenamefont
  {Song}}]{jo_enhancement_2021}%
  \BibitemOpen
  \bibfield  {author} {\bibinfo {author} {\bibfnamefont {M.-k.}\ \bibnamefont
  {Jo}}, \bibinfo {author} {\bibfnamefont {H.}~\bibnamefont {Heo}}, \bibinfo
  {author} {\bibfnamefont {J.-H.}\ \bibnamefont {Lee}}, \bibinfo {author}
  {\bibfnamefont {S.}~\bibnamefont {Choi}}, \bibinfo {author} {\bibfnamefont
  {A.}~\bibnamefont {Kim}}, \bibinfo {author} {\bibfnamefont {H.~B.}\
  \bibnamefont {Jeong}}, \bibinfo {author} {\bibfnamefont {H.~Y.}\ \bibnamefont
  {Jeong}}, \bibinfo {author} {\bibfnamefont {J.~M.}\ \bibnamefont {Yuk}},
  \bibinfo {author} {\bibfnamefont {D.}~\bibnamefont {Eom}}, \bibinfo {author}
  {\bibfnamefont {J.}~\bibnamefont {Jahng}}, \bibinfo {author} {\bibfnamefont
  {E.~S.}\ \bibnamefont {Lee}}, \bibinfo {author} {\bibfnamefont {I.-y.}\
  \bibnamefont {Jung}}, \bibinfo {author} {\bibfnamefont {S.~R.}\ \bibnamefont
  {Cho}}, \bibinfo {author} {\bibfnamefont {J.}~\bibnamefont {Kim}}, \bibinfo
  {author} {\bibfnamefont {S.}~\bibnamefont {Cho}}, \bibinfo {author}
  {\bibfnamefont {K.}~\bibnamefont {Kang}},\ and\ \bibinfo {author}
  {\bibfnamefont {S.}~\bibnamefont {Song}},\ }\href
  {https://doi.org/10.1021/acsnano.1c06752} {\bibfield  {journal} {\bibinfo
  {journal} {ACS Nano}\ }\textbf {\bibinfo {volume} {15}},\ \bibinfo {pages}
  {18113} (\bibinfo {year} {2021})},\ \bibinfo {note} {publisher: American
  Chemical Society}\BibitemShut {NoStop}%
\bibitem [{\citenamefont {Zhou}\ \emph {et~al.}(2016)\citenamefont {Zhou},
  \citenamefont {Li}, \citenamefont {Waugh}, \citenamefont {Parham},
  \citenamefont {Kim}, \citenamefont {Sears}, \citenamefont {Gomes},
  \citenamefont {Kee}, \citenamefont {Kim},\ and\ \citenamefont
  {Dessau}}]{zhou_angle-resolved_2016}%
  \BibitemOpen
  \bibfield  {author} {\bibinfo {author} {\bibfnamefont {X.}~\bibnamefont
  {Zhou}}, \bibinfo {author} {\bibfnamefont {H.}~\bibnamefont {Li}}, \bibinfo
  {author} {\bibfnamefont {J.~A.}\ \bibnamefont {Waugh}}, \bibinfo {author}
  {\bibfnamefont {S.}~\bibnamefont {Parham}}, \bibinfo {author} {\bibfnamefont
  {H.-S.}\ \bibnamefont {Kim}}, \bibinfo {author} {\bibfnamefont {J.~A.}\
  \bibnamefont {Sears}}, \bibinfo {author} {\bibfnamefont {A.}~\bibnamefont
  {Gomes}}, \bibinfo {author} {\bibfnamefont {H.-Y.}\ \bibnamefont {Kee}},
  \bibinfo {author} {\bibfnamefont {Y.-J.}\ \bibnamefont {Kim}},\ and\ \bibinfo
  {author} {\bibfnamefont {D.~S.}\ \bibnamefont {Dessau}},\ }\href
  {https://doi.org/10.1103/PhysRevB.94.161106} {\bibfield  {journal} {\bibinfo
  {journal} {Physical Review B}\ }\textbf {\bibinfo {volume} {94}},\ \bibinfo
  {pages} {161106(R)} (\bibinfo {year} {2016})},\ \bibinfo {note} {publisher:
  American Physical Society}\BibitemShut {NoStop}%
\bibitem [{\citenamefont {Zhou}\ \emph {et~al.}(2019)\citenamefont {Zhou},
  \citenamefont {Balgley}, \citenamefont {Lampen-Kelley}, \citenamefont {Yan},
  \citenamefont {Mandrus},\ and\ \citenamefont
  {Henriksen}}]{zhou_evidence_2019}%
  \BibitemOpen
  \bibfield  {author} {\bibinfo {author} {\bibfnamefont {B.}~\bibnamefont
  {Zhou}}, \bibinfo {author} {\bibfnamefont {J.}~\bibnamefont {Balgley}},
  \bibinfo {author} {\bibfnamefont {P.}~\bibnamefont {Lampen-Kelley}}, \bibinfo
  {author} {\bibfnamefont {J.-Q.}\ \bibnamefont {Yan}}, \bibinfo {author}
  {\bibfnamefont {D.~G.}\ \bibnamefont {Mandrus}},\ and\ \bibinfo {author}
  {\bibfnamefont {E.~A.}\ \bibnamefont {Henriksen}},\ }\href
  {https://doi.org/10.1103/PhysRevB.100.165426} {\bibfield  {journal} {\bibinfo
   {journal} {Physical Review B}\ }\textbf {\bibinfo {volume} {100}},\ \bibinfo
  {pages} {165426} (\bibinfo {year} {2019})},\ \bibinfo {note} {publisher:
  American Physical Society}\BibitemShut {NoStop}%
\bibitem [{\citenamefont {Rizzo}\ \emph {et~al.}(2020)\citenamefont {Rizzo},
  \citenamefont {Jessen}, \citenamefont {Sun}, \citenamefont {Ruta},
  \citenamefont {Zhang}, \citenamefont {Yan}, \citenamefont {Xian},
  \citenamefont {McLeod}, \citenamefont {Berkowitz}, \citenamefont {Watanabe},
  \citenamefont {Taniguchi}, \citenamefont {Nagler}, \citenamefont {Mandrus},
  \citenamefont {Rubio}, \citenamefont {Fogler}, \citenamefont {Millis},
  \citenamefont {Hone}, \citenamefont {Dean},\ and\ \citenamefont
  {Basov}}]{rizzo_charge-transfer_2020}%
  \BibitemOpen
  \bibfield  {author} {\bibinfo {author} {\bibfnamefont {D.~J.}\ \bibnamefont
  {Rizzo}}, \bibinfo {author} {\bibfnamefont {B.~S.}\ \bibnamefont {Jessen}},
  \bibinfo {author} {\bibfnamefont {Z.}~\bibnamefont {Sun}}, \bibinfo {author}
  {\bibfnamefont {F.~L.}\ \bibnamefont {Ruta}}, \bibinfo {author}
  {\bibfnamefont {J.}~\bibnamefont {Zhang}}, \bibinfo {author} {\bibfnamefont
  {J.-Q.}\ \bibnamefont {Yan}}, \bibinfo {author} {\bibfnamefont
  {L.}~\bibnamefont {Xian}}, \bibinfo {author} {\bibfnamefont {A.~S.}\
  \bibnamefont {McLeod}}, \bibinfo {author} {\bibfnamefont {M.~E.}\
  \bibnamefont {Berkowitz}}, \bibinfo {author} {\bibfnamefont {K.}~\bibnamefont
  {Watanabe}}, \bibinfo {author} {\bibfnamefont {T.}~\bibnamefont {Taniguchi}},
  \bibinfo {author} {\bibfnamefont {S.~E.}\ \bibnamefont {Nagler}}, \bibinfo
  {author} {\bibfnamefont {D.~G.}\ \bibnamefont {Mandrus}}, \bibinfo {author}
  {\bibfnamefont {A.}~\bibnamefont {Rubio}}, \bibinfo {author} {\bibfnamefont
  {M.~M.}\ \bibnamefont {Fogler}}, \bibinfo {author} {\bibfnamefont {A.~J.}\
  \bibnamefont {Millis}}, \bibinfo {author} {\bibfnamefont {J.~C.}\
  \bibnamefont {Hone}}, \bibinfo {author} {\bibfnamefont {C.~R.}\ \bibnamefont
  {Dean}},\ and\ \bibinfo {author} {\bibfnamefont {D.~N.}\ \bibnamefont
  {Basov}},\ }\href {https://doi.org/10.1021/acs.nanolett.0c03466} {\bibfield
  {journal} {\bibinfo  {journal} {Nano Letters}\ }\textbf {\bibinfo {volume}
  {20}},\ \bibinfo {pages} {8438} (\bibinfo {year} {2020})},\ \bibinfo {note}
  {publisher: American Chemical Society}\BibitemShut {NoStop}%
\bibitem [{\citenamefont {Biswas}\ \emph {et~al.}(2019)\citenamefont {Biswas},
  \citenamefont {Li}, \citenamefont {Winter}, \citenamefont {Knolle},\ and\
  \citenamefont {Valent\'{\i}}}]{biswas_electronic_2019}%
  \BibitemOpen
  \bibfield  {author} {\bibinfo {author} {\bibfnamefont {S.}~\bibnamefont
  {Biswas}}, \bibinfo {author} {\bibfnamefont {Y.}~\bibnamefont {Li}}, \bibinfo
  {author} {\bibfnamefont {S.~M.}\ \bibnamefont {Winter}}, \bibinfo {author}
  {\bibfnamefont {J.}~\bibnamefont {Knolle}},\ and\ \bibinfo {author}
  {\bibfnamefont {R.}~\bibnamefont {Valent\'{\i}}},\ }\href
  {https://doi.org/10.1103/PhysRevLett.123.237201} {\bibfield  {journal}
  {\bibinfo  {journal} {Physical Review Letters}\ }\textbf {\bibinfo {volume}
  {123}},\ \bibinfo {pages} {237201} (\bibinfo {year} {2019})},\ \bibinfo
  {note} {publisher: American Physical Society}\BibitemShut {NoStop}%
\bibitem [{\citenamefont {Mashhadi}\ \emph {et~al.}(2019)\citenamefont
  {Mashhadi}, \citenamefont {Kim}, \citenamefont {Kim}, \citenamefont {Weber},
  \citenamefont {Taniguchi}, \citenamefont {Watanabe}, \citenamefont {Park},
  \citenamefont {Lotsch}, \citenamefont {Smet}, \citenamefont {Burghard},\ and\
  \citenamefont {Kern}}]{mashhadi_spin-split_2019}%
  \BibitemOpen
  \bibfield  {author} {\bibinfo {author} {\bibfnamefont {S.}~\bibnamefont
  {Mashhadi}}, \bibinfo {author} {\bibfnamefont {Y.}~\bibnamefont {Kim}},
  \bibinfo {author} {\bibfnamefont {J.}~\bibnamefont {Kim}}, \bibinfo {author}
  {\bibfnamefont {D.}~\bibnamefont {Weber}}, \bibinfo {author} {\bibfnamefont
  {T.}~\bibnamefont {Taniguchi}}, \bibinfo {author} {\bibfnamefont
  {K.}~\bibnamefont {Watanabe}}, \bibinfo {author} {\bibfnamefont
  {N.}~\bibnamefont {Park}}, \bibinfo {author} {\bibfnamefont {B.}~\bibnamefont
  {Lotsch}}, \bibinfo {author} {\bibfnamefont {J.~H.}\ \bibnamefont {Smet}},
  \bibinfo {author} {\bibfnamefont {M.}~\bibnamefont {Burghard}},\ and\
  \bibinfo {author} {\bibfnamefont {K.}~\bibnamefont {Kern}},\ }\href
  {https://doi.org/10.1021/acs.nanolett.9b01691} {\bibfield  {journal}
  {\bibinfo  {journal} {Nano Letters}\ }\textbf {\bibinfo {volume} {19}},\
  \bibinfo {pages} {4659} (\bibinfo {year} {2019})},\ \bibinfo {note}
  {publisher: American Chemical Society}\BibitemShut {NoStop}%
\bibitem [{\citenamefont {Gerber}\ \emph {et~al.}(2020)\citenamefont {Gerber},
  \citenamefont {Yao}, \citenamefont {Arias},\ and\ \citenamefont
  {Kim}}]{gerber_ab_2020}%
  \BibitemOpen
  \bibfield  {author} {\bibinfo {author} {\bibfnamefont {E.}~\bibnamefont
  {Gerber}}, \bibinfo {author} {\bibfnamefont {Y.}~\bibnamefont {Yao}},
  \bibinfo {author} {\bibfnamefont {T.~A.}\ \bibnamefont {Arias}},\ and\
  \bibinfo {author} {\bibfnamefont {E.-A.}\ \bibnamefont {Kim}},\ }\href
  {https://doi.org/10.1103/PhysRevLett.124.106804} {\bibfield  {journal}
  {\bibinfo  {journal} {Physical Review Letters}\ }\textbf {\bibinfo {volume}
  {124}},\ \bibinfo {pages} {106804} (\bibinfo {year} {2020})},\ \bibinfo
  {note} {publisher: American Physical Society}\BibitemShut {NoStop}%
\bibitem [{\citenamefont {Souza}\ \emph {et~al.}(2022)\citenamefont {Souza},
  \citenamefont {Deus}, \citenamefont {Brito},\ and\ \citenamefont
  {Miwa}}]{souza_magnetic_2022}%
  \BibitemOpen
  \bibfield  {author} {\bibinfo {author} {\bibfnamefont {P.~H.}\ \bibnamefont
  {Souza}}, \bibinfo {author} {\bibfnamefont {D.~P. d.~A.}\ \bibnamefont
  {Deus}}, \bibinfo {author} {\bibfnamefont {W.~H.}\ \bibnamefont {Brito}},\
  and\ \bibinfo {author} {\bibfnamefont {R.~H.}\ \bibnamefont {Miwa}},\ }\href
  {https://doi.org/10.1103/PhysRevB.106.155118} {\bibfield  {journal} {\bibinfo
   {journal} {Physical Review B}\ }\textbf {\bibinfo {volume} {106}},\ \bibinfo
  {pages} {155118} (\bibinfo {year} {2022})},\ \bibinfo {note} {publisher:
  American Physical Society}\BibitemShut {NoStop}%
\bibitem [{\citenamefont {Zheng}\ \emph {et~al.}(2023)\citenamefont {Zheng},
  \citenamefont {Jia}, \citenamefont {Ren}, \citenamefont {Yang}, \citenamefont
  {Wu}, \citenamefont {Shi}, \citenamefont {Tanigaki},\ and\ \citenamefont
  {Du}}]{zheng_tunneling_2023}%
  \BibitemOpen
  \bibfield  {author} {\bibinfo {author} {\bibfnamefont {X.}~\bibnamefont
  {Zheng}}, \bibinfo {author} {\bibfnamefont {K.}~\bibnamefont {Jia}}, \bibinfo
  {author} {\bibfnamefont {J.}~\bibnamefont {Ren}}, \bibinfo {author}
  {\bibfnamefont {C.}~\bibnamefont {Yang}}, \bibinfo {author} {\bibfnamefont
  {X.}~\bibnamefont {Wu}}, \bibinfo {author} {\bibfnamefont {Y.}~\bibnamefont
  {Shi}}, \bibinfo {author} {\bibfnamefont {K.}~\bibnamefont {Tanigaki}},\ and\
  \bibinfo {author} {\bibfnamefont {R.-R.}\ \bibnamefont {Du}},\ }\href
  {https://doi.org/10.1103/PhysRevB.107.195107} {\bibfield  {journal} {\bibinfo
   {journal} {Physical Review B}\ }\textbf {\bibinfo {volume} {107}},\ \bibinfo
  {pages} {195107} (\bibinfo {year} {2023})},\ \bibinfo {note} {publisher:
  American Physical Society}\BibitemShut {NoStop}%
\bibitem [{\citenamefont {Iyikanat}\ \emph {et~al.}(2018)\citenamefont
  {Iyikanat}, \citenamefont {Yagmurcukardes}, \citenamefont {Senger},\ and\
  \citenamefont {Sahin}}]{iyikanat_tuning_2018}%
  \BibitemOpen
  \bibfield  {author} {\bibinfo {author} {\bibfnamefont {F.}~\bibnamefont
  {Iyikanat}}, \bibinfo {author} {\bibfnamefont {M.}~\bibnamefont
  {Yagmurcukardes}}, \bibinfo {author} {\bibfnamefont {R.~T.}\ \bibnamefont
  {Senger}},\ and\ \bibinfo {author} {\bibfnamefont {H.}~\bibnamefont
  {Sahin}},\ }\href {https://doi.org/10.1039/C7TC05266A} {\bibfield  {journal}
  {\bibinfo  {journal} {Journal of Materials Chemistry C}\ }\textbf {\bibinfo
  {volume} {6}},\ \bibinfo {pages} {2019} (\bibinfo {year} {2018})},\ \bibinfo
  {note} {publisher: The Royal Society of Chemistry}\BibitemShut {NoStop}%
\bibitem [{\citenamefont {Kaib}\ \emph {et~al.}(2021)\citenamefont {Kaib},
  \citenamefont {Biswas}, \citenamefont {Riedl}, \citenamefont {Winter},\ and\
  \citenamefont {Valent\'{\i}}}]{kaib_magnetoelastic_2021}%
  \BibitemOpen
  \bibfield  {author} {\bibinfo {author} {\bibfnamefont {D.~A.~S.}\
  \bibnamefont {Kaib}}, \bibinfo {author} {\bibfnamefont {S.}~\bibnamefont
  {Biswas}}, \bibinfo {author} {\bibfnamefont {K.}~\bibnamefont {Riedl}},
  \bibinfo {author} {\bibfnamefont {S.~M.}\ \bibnamefont {Winter}},\ and\
  \bibinfo {author} {\bibfnamefont {R.}~\bibnamefont {Valent\'{\i}}},\ }\href
  {https://doi.org/10.1103/PhysRevB.103.L140402} {\bibfield  {journal}
  {\bibinfo  {journal} {Physical Review B}\ }\textbf {\bibinfo {volume}
  {103}},\ \bibinfo {pages} {L140402} (\bibinfo {year} {2021})},\ \bibinfo
  {note} {publisher: American Physical Society}\BibitemShut {NoStop}%
\bibitem [{\citenamefont {Plumb}\ \emph {et~al.}(2014)\citenamefont {Plumb},
  \citenamefont {Clancy}, \citenamefont {Sandilands}, \citenamefont {Shankar},
  \citenamefont {Hu}, \citenamefont {Burch}, \citenamefont {Kee},\ and\
  \citenamefont {Kim}}]{plumb_ensuremathalphaensuremath-mathrmrucl_3_2014}%
  \BibitemOpen
  \bibfield  {author} {\bibinfo {author} {\bibfnamefont {K.~W.}\ \bibnamefont
  {Plumb}}, \bibinfo {author} {\bibfnamefont {J.~P.}\ \bibnamefont {Clancy}},
  \bibinfo {author} {\bibfnamefont {L.~J.}\ \bibnamefont {Sandilands}},
  \bibinfo {author} {\bibfnamefont {V.~V.}\ \bibnamefont {Shankar}}, \bibinfo
  {author} {\bibfnamefont {Y.~F.}\ \bibnamefont {Hu}}, \bibinfo {author}
  {\bibfnamefont {K.~S.}\ \bibnamefont {Burch}}, \bibinfo {author}
  {\bibfnamefont {H.-Y.}\ \bibnamefont {Kee}},\ and\ \bibinfo {author}
  {\bibfnamefont {Y.-J.}\ \bibnamefont {Kim}},\ }\href
  {https://doi.org/10.1103/PhysRevB.90.041112} {\bibfield  {journal} {\bibinfo
  {journal} {Physical Review B}\ }\textbf {\bibinfo {volume} {90}},\ \bibinfo
  {pages} {041112(R)} (\bibinfo {year} {2014})},\ \bibinfo {note} {publisher:
  American Physical Society}\BibitemShut {NoStop}%
\bibitem [{\citenamefont {Lu}\ \emph {et~al.}(2023)\citenamefont {Lu},
  \citenamefont {Lee}, \citenamefont {Cha}, \citenamefont {Zhang},\ and\
  \citenamefont {Chung}}]{lu_electronic_2023}%
  \BibitemOpen
  \bibfield  {author} {\bibinfo {author} {\bibfnamefont {W.}~\bibnamefont
  {Lu}}, \bibinfo {author} {\bibfnamefont {H.}~\bibnamefont {Lee}}, \bibinfo
  {author} {\bibfnamefont {J.}~\bibnamefont {Cha}}, \bibinfo {author}
  {\bibfnamefont {J.}~\bibnamefont {Zhang}},\ and\ \bibinfo {author}
  {\bibfnamefont {I.}~\bibnamefont {Chung}},\ }\href
  {https://doi.org/10.1002/ange.202219344} {\bibfield  {journal} {\bibinfo
  {journal} {Angewandte Chemie}\ }\textbf {\bibinfo {volume} {135}},\ \bibinfo
  {pages} {e202219344} (\bibinfo {year} {2023})},\ \bibinfo {note} {publisher:
  Wiley Onlin Library}\BibitemShut {NoStop}%
\bibitem [{\citenamefont {Koitzsch}\ \emph {et~al.}(2016)\citenamefont
  {Koitzsch}, \citenamefont {Habenicht}, \citenamefont {M\"uller},
  \citenamefont {Knupfer}, \citenamefont {B\"uchner}, \citenamefont {Kandpal},
  \citenamefont {van~den Brink}, \citenamefont {Nowak}, \citenamefont
  {Isaeva},\ and\ \citenamefont {Doert}}]{koitzsch_j_eff_2016}%
  \BibitemOpen
  \bibfield  {author} {\bibinfo {author} {\bibfnamefont {A.}~\bibnamefont
  {Koitzsch}}, \bibinfo {author} {\bibfnamefont {C.}~\bibnamefont {Habenicht}},
  \bibinfo {author} {\bibfnamefont {E.}~\bibnamefont {M\"uller}}, \bibinfo
  {author} {\bibfnamefont {M.}~\bibnamefont {Knupfer}}, \bibinfo {author}
  {\bibfnamefont {B.}~\bibnamefont {B\"uchner}}, \bibinfo {author}
  {\bibfnamefont {H.~C.}\ \bibnamefont {Kandpal}}, \bibinfo {author}
  {\bibfnamefont {J.}~\bibnamefont {van~den Brink}}, \bibinfo {author}
  {\bibfnamefont {D.}~\bibnamefont {Nowak}}, \bibinfo {author} {\bibfnamefont
  {A.}~\bibnamefont {Isaeva}},\ and\ \bibinfo {author} {\bibfnamefont
  {T.}~\bibnamefont {Doert}},\ }\href
  {https://doi.org/10.1103/PhysRevLett.117.126403} {\bibfield  {journal}
  {\bibinfo  {journal} {Phys. Rev. Lett.}\ }\textbf {\bibinfo {volume} {117}},\
  \bibinfo {pages} {126403} (\bibinfo {year} {2016})}\BibitemShut {NoStop}%
\bibitem [{\citenamefont {Jackeli}\ and\ \citenamefont
  {Khaliullin}(2009)}]{jackeli_mott_2009}%
  \BibitemOpen
  \bibfield  {author} {\bibinfo {author} {\bibfnamefont {G.}~\bibnamefont
  {Jackeli}}\ and\ \bibinfo {author} {\bibfnamefont {G.}~\bibnamefont
  {Khaliullin}},\ }\href {https://doi.org/10.1103/PhysRevLett.102.017205}
  {\bibfield  {journal} {\bibinfo  {journal} {Physical Review Letters}\
  }\textbf {\bibinfo {volume} {102}},\ \bibinfo {pages} {017205} (\bibinfo
  {year} {2009})},\ \bibinfo {note} {publisher: American Physical
  Society}\BibitemShut {NoStop}%
\bibitem [{\citenamefont {Sinn}\ \emph {et~al.}(2016)\citenamefont {Sinn},
  \citenamefont {Kim}, \citenamefont {Kim}, \citenamefont {Lee}, \citenamefont
  {Won}, \citenamefont {Oh}, \citenamefont {Han}, \citenamefont {Chang},
  \citenamefont {Hur}, \citenamefont {Sato}, \citenamefont {Park},
  \citenamefont {Kim}, \citenamefont {Kim},\ and\ \citenamefont
  {Noh}}]{sinn_electronic_2016}%
  \BibitemOpen
  \bibfield  {author} {\bibinfo {author} {\bibfnamefont {S.}~\bibnamefont
  {Sinn}}, \bibinfo {author} {\bibfnamefont {C.~H.}\ \bibnamefont {Kim}},
  \bibinfo {author} {\bibfnamefont {B.~H.}\ \bibnamefont {Kim}}, \bibinfo
  {author} {\bibfnamefont {K.~D.}\ \bibnamefont {Lee}}, \bibinfo {author}
  {\bibfnamefont {C.~J.}\ \bibnamefont {Won}}, \bibinfo {author} {\bibfnamefont
  {J.~S.}\ \bibnamefont {Oh}}, \bibinfo {author} {\bibfnamefont
  {M.}~\bibnamefont {Han}}, \bibinfo {author} {\bibfnamefont {Y.~J.}\
  \bibnamefont {Chang}}, \bibinfo {author} {\bibfnamefont {N.}~\bibnamefont
  {Hur}}, \bibinfo {author} {\bibfnamefont {H.}~\bibnamefont {Sato}}, \bibinfo
  {author} {\bibfnamefont {B.-G.}\ \bibnamefont {Park}}, \bibinfo {author}
  {\bibfnamefont {C.}~\bibnamefont {Kim}}, \bibinfo {author} {\bibfnamefont
  {H.-D.}\ \bibnamefont {Kim}},\ and\ \bibinfo {author} {\bibfnamefont {T.~W.}\
  \bibnamefont {Noh}},\ }\href {https://doi.org/10.1038/srep39544} {\bibfield
  {journal} {\bibinfo  {journal} {Scientific Reports}\ }\textbf {\bibinfo
  {volume} {6}},\ \bibinfo {pages} {39544} (\bibinfo {year} {2016})},\ \bibinfo
  {note} {number: 1 Publisher: Nature Publishing Group}\BibitemShut {NoStop}%
\bibitem [{\citenamefont {Bu}\ \emph {et~al.}(2019)\citenamefont {Bu},
  \citenamefont {Zhang}, \citenamefont {Fei}, \citenamefont {Wu}, \citenamefont
  {Zheng}, \citenamefont {Gao}, \citenamefont {Luo}, \citenamefont {Sun},\ and\
  \citenamefont {Yin}}]{bu_possible_2019}%
  \BibitemOpen
  \bibfield  {author} {\bibinfo {author} {\bibfnamefont {K.}~\bibnamefont
  {Bu}}, \bibinfo {author} {\bibfnamefont {W.}~\bibnamefont {Zhang}}, \bibinfo
  {author} {\bibfnamefont {Y.}~\bibnamefont {Fei}}, \bibinfo {author}
  {\bibfnamefont {Z.}~\bibnamefont {Wu}}, \bibinfo {author} {\bibfnamefont
  {Y.}~\bibnamefont {Zheng}}, \bibinfo {author} {\bibfnamefont
  {J.}~\bibnamefont {Gao}}, \bibinfo {author} {\bibfnamefont {X.}~\bibnamefont
  {Luo}}, \bibinfo {author} {\bibfnamefont {Y.-P.}\ \bibnamefont {Sun}},\ and\
  \bibinfo {author} {\bibfnamefont {Y.}~\bibnamefont {Yin}},\ }\href
  {https://doi.org/10.1038/s42005-019-0247-0} {\bibfield  {journal} {\bibinfo
  {journal} {Communications Physics}\ }\textbf {\bibinfo {volume} {2}},\
  \bibinfo {pages} {1} (\bibinfo {year} {2019})},\ \bibinfo {note} {number: 1
  Publisher: Nature Publishing Group}\BibitemShut {NoStop}%
\bibitem [{\citenamefont {Dymkowski}\ and\ \citenamefont
  {Ederer}(2014)}]{dymkowski_strain-induced_2014}%
  \BibitemOpen
  \bibfield  {author} {\bibinfo {author} {\bibfnamefont {K.}~\bibnamefont
  {Dymkowski}}\ and\ \bibinfo {author} {\bibfnamefont {C.}~\bibnamefont
  {Ederer}},\ }\href {https://doi.org/10.1103/PhysRevB.89.161109} {\bibfield
  {journal} {\bibinfo  {journal} {Physical Review B}\ }\textbf {\bibinfo
  {volume} {89}},\ \bibinfo {pages} {161109(R)} (\bibinfo {year} {2014})},\
  \bibinfo {note} {publisher: American Physical Society}\BibitemShut {NoStop}%
\bibitem [{\citenamefont {Vatansever}\ \emph {et~al.}(2019)\citenamefont
  {Vatansever}, \citenamefont {Sarikurt}, \citenamefont {Ersan}, \citenamefont
  {Kadioglu}, \citenamefont {Üzengi Aktürk}, \citenamefont {Yüksel},
  \citenamefont {Ataca}, \citenamefont {Aktürk},\ and\ \citenamefont
  {Akinci}}]{vatansever_strain_2019}%
  \BibitemOpen
  \bibfield  {author} {\bibinfo {author} {\bibfnamefont {E.}~\bibnamefont
  {Vatansever}}, \bibinfo {author} {\bibfnamefont {S.}~\bibnamefont
  {Sarikurt}}, \bibinfo {author} {\bibfnamefont {F.}~\bibnamefont {Ersan}},
  \bibinfo {author} {\bibfnamefont {Y.}~\bibnamefont {Kadioglu}}, \bibinfo
  {author} {\bibfnamefont {O.}~\bibnamefont {Üzengi Aktürk}}, \bibinfo
  {author} {\bibfnamefont {Y.}~\bibnamefont {Yüksel}}, \bibinfo {author}
  {\bibfnamefont {C.}~\bibnamefont {Ataca}}, \bibinfo {author} {\bibfnamefont
  {E.}~\bibnamefont {Aktürk}},\ and\ \bibinfo {author} {\bibfnamefont
  {U.}~\bibnamefont {Akinci}},\ }\href {https://doi.org/10.1063/1.5078713}
  {\bibfield  {journal} {\bibinfo  {journal} {Journal of Applied Physics}\
  }\textbf {\bibinfo {volume} {125}},\ \bibinfo {pages} {083903} (\bibinfo
  {year} {2019})}\BibitemShut {NoStop}%
\bibitem [{\citenamefont {Wang}\ \emph {et~al.}(2022)\citenamefont {Wang},
  \citenamefont {Liu}, \citenamefont {Zhao}, \citenamefont {Zheng},
  \citenamefont {Yang}, \citenamefont {Wang}, \citenamefont {Yang},
  \citenamefont {Wu},\ and\ \citenamefont {Gao}}]{wang_identification_2022}%
  \BibitemOpen
  \bibfield  {author} {\bibinfo {author} {\bibfnamefont {Z.}~\bibnamefont
  {Wang}}, \bibinfo {author} {\bibfnamefont {L.}~\bibnamefont {Liu}}, \bibinfo
  {author} {\bibfnamefont {M.}~\bibnamefont {Zhao}}, \bibinfo {author}
  {\bibfnamefont {H.}~\bibnamefont {Zheng}}, \bibinfo {author} {\bibfnamefont
  {K.}~\bibnamefont {Yang}}, \bibinfo {author} {\bibfnamefont {C.}~\bibnamefont
  {Wang}}, \bibinfo {author} {\bibfnamefont {F.}~\bibnamefont {Yang}}, \bibinfo
  {author} {\bibfnamefont {H.}~\bibnamefont {Wu}},\ and\ \bibinfo {author}
  {\bibfnamefont {C.}~\bibnamefont {Gao}},\ }\href
  {https://doi.org/10.1007/s44214-022-00016-8} {\bibfield  {journal} {\bibinfo
  {journal} {Quantum Frontiers}\ }\textbf {\bibinfo {volume} {1}},\ \bibinfo
  {pages} {16} (\bibinfo {year} {2022})}\BibitemShut {NoStop}%
\bibitem [{\citenamefont {Hÿtch}\ \emph {et~al.}(1998)\citenamefont {Hÿtch},
  \citenamefont {Snoeck},\ and\ \citenamefont
  {Kilaas}}]{hytch_quantitative_1998}%
  \BibitemOpen
  \bibfield  {author} {\bibinfo {author} {\bibfnamefont {M.~J.}\ \bibnamefont
  {Hÿtch}}, \bibinfo {author} {\bibfnamefont {E.}~\bibnamefont {Snoeck}},\
  and\ \bibinfo {author} {\bibfnamefont {R.}~\bibnamefont {Kilaas}},\ }\href
  {https://doi.org/https://doi.org/10.1016/S0304-3991(98)00035-7} {\bibfield
  {journal} {\bibinfo  {journal} {Ultramicroscopy}\ }\textbf {\bibinfo {volume}
  {74}},\ \bibinfo {pages} {131} (\bibinfo {year} {1998})}\BibitemShut
  {NoStop}%
\bibitem [{\citenamefont {Cai}\ \emph {et~al.}(2016)\citenamefont {Cai},
  \citenamefont {Ruan}, \citenamefont {Peng}, \citenamefont {Ye}, \citenamefont
  {Li}, \citenamefont {Hao}, \citenamefont {Zhou}, \citenamefont {Lee},\ and\
  \citenamefont {Wang}}]{cai_visualizing_2016}%
  \BibitemOpen
  \bibfield  {author} {\bibinfo {author} {\bibfnamefont {P.}~\bibnamefont
  {Cai}}, \bibinfo {author} {\bibfnamefont {W.}~\bibnamefont {Ruan}}, \bibinfo
  {author} {\bibfnamefont {Y.}~\bibnamefont {Peng}}, \bibinfo {author}
  {\bibfnamefont {C.}~\bibnamefont {Ye}}, \bibinfo {author} {\bibfnamefont
  {X.}~\bibnamefont {Li}}, \bibinfo {author} {\bibfnamefont {Z.}~\bibnamefont
  {Hao}}, \bibinfo {author} {\bibfnamefont {X.}~\bibnamefont {Zhou}}, \bibinfo
  {author} {\bibfnamefont {D.-H.}\ \bibnamefont {Lee}},\ and\ \bibinfo {author}
  {\bibfnamefont {Y.}~\bibnamefont {Wang}},\ }\href
  {https://doi.org/10.1038/nphys3840} {\bibfield  {journal} {\bibinfo
  {journal} {Nature Physics}\ }\textbf {\bibinfo {volume} {12}},\ \bibinfo
  {pages} {1047} (\bibinfo {year} {2016})},\ \bibinfo {note} {number: 11
  Publisher: Nature Publishing Group}\BibitemShut {NoStop}%
\bibitem [{\citenamefont {Battisti}\ \emph {et~al.}(2017)\citenamefont
  {Battisti}, \citenamefont {Bastiaans}, \citenamefont {Fedoseev},
  \citenamefont {de~la Torre}, \citenamefont {Iliopoulos}, \citenamefont
  {Tamai}, \citenamefont {Hunter}, \citenamefont {Perry}, \citenamefont
  {Zaanen}, \citenamefont {Baumberger},\ and\ \citenamefont
  {Allan}}]{battisti_universality_2017}%
  \BibitemOpen
  \bibfield  {author} {\bibinfo {author} {\bibfnamefont {I.}~\bibnamefont
  {Battisti}}, \bibinfo {author} {\bibfnamefont {K.~M.}\ \bibnamefont
  {Bastiaans}}, \bibinfo {author} {\bibfnamefont {V.}~\bibnamefont {Fedoseev}},
  \bibinfo {author} {\bibfnamefont {A.}~\bibnamefont {de~la Torre}}, \bibinfo
  {author} {\bibfnamefont {N.}~\bibnamefont {Iliopoulos}}, \bibinfo {author}
  {\bibfnamefont {A.}~\bibnamefont {Tamai}}, \bibinfo {author} {\bibfnamefont
  {E.~C.}\ \bibnamefont {Hunter}}, \bibinfo {author} {\bibfnamefont {R.~S.}\
  \bibnamefont {Perry}}, \bibinfo {author} {\bibfnamefont {J.}~\bibnamefont
  {Zaanen}}, \bibinfo {author} {\bibfnamefont {F.}~\bibnamefont {Baumberger}},\
  and\ \bibinfo {author} {\bibfnamefont {M.~P.}\ \bibnamefont {Allan}},\ }\href
  {https://doi.org/10.1038/nphys3894} {\bibfield  {journal} {\bibinfo
  {journal} {Nature Physics}\ }\textbf {\bibinfo {volume} {13}},\ \bibinfo
  {pages} {21} (\bibinfo {year} {2017})},\ \bibinfo {note} {number: 1
  Publisher: Nature Publishing Group}\BibitemShut {NoStop}%
\bibitem [{\citenamefont {da~Silva~Neto}\ \emph {et~al.}(2016)\citenamefont
  {da~Silva~Neto}, \citenamefont {Yu}, \citenamefont {Minola}, \citenamefont
  {Sutarto}, \citenamefont {Schierle}, \citenamefont {Boschini}, \citenamefont
  {Zonno}, \citenamefont {Bluschke}, \citenamefont {Higgins}, \citenamefont
  {Li}, \citenamefont {Yu}, \citenamefont {Weschke}, \citenamefont {He},
  \citenamefont {Le~Tacon}, \citenamefont {Greene}, \citenamefont {Greven},
  \citenamefont {Sawatzky}, \citenamefont {Keimer},\ and\ \citenamefont
  {Damascelli}}]{da_silva_neto_doping-dependent_2016}%
  \BibitemOpen
  \bibfield  {author} {\bibinfo {author} {\bibfnamefont {E.~H.}\ \bibnamefont
  {da~Silva~Neto}}, \bibinfo {author} {\bibfnamefont {B.}~\bibnamefont {Yu}},
  \bibinfo {author} {\bibfnamefont {M.}~\bibnamefont {Minola}}, \bibinfo
  {author} {\bibfnamefont {R.}~\bibnamefont {Sutarto}}, \bibinfo {author}
  {\bibfnamefont {E.}~\bibnamefont {Schierle}}, \bibinfo {author}
  {\bibfnamefont {F.}~\bibnamefont {Boschini}}, \bibinfo {author}
  {\bibfnamefont {M.}~\bibnamefont {Zonno}}, \bibinfo {author} {\bibfnamefont
  {M.}~\bibnamefont {Bluschke}}, \bibinfo {author} {\bibfnamefont
  {J.}~\bibnamefont {Higgins}}, \bibinfo {author} {\bibfnamefont
  {Y.}~\bibnamefont {Li}}, \bibinfo {author} {\bibfnamefont {G.}~\bibnamefont
  {Yu}}, \bibinfo {author} {\bibfnamefont {E.}~\bibnamefont {Weschke}},
  \bibinfo {author} {\bibfnamefont {F.}~\bibnamefont {He}}, \bibinfo {author}
  {\bibfnamefont {M.}~\bibnamefont {Le~Tacon}}, \bibinfo {author}
  {\bibfnamefont {R.~L.}\ \bibnamefont {Greene}}, \bibinfo {author}
  {\bibfnamefont {M.}~\bibnamefont {Greven}}, \bibinfo {author} {\bibfnamefont
  {G.~A.}\ \bibnamefont {Sawatzky}}, \bibinfo {author} {\bibfnamefont
  {B.}~\bibnamefont {Keimer}},\ and\ \bibinfo {author} {\bibfnamefont
  {A.}~\bibnamefont {Damascelli}},\ }\href
  {https://doi.org/10.1126/sciadv.1600782} {\bibfield  {journal} {\bibinfo
  {journal} {Science advances}\ }\textbf {\bibinfo {volume} {2}},\ \bibinfo
  {pages} {e1600782} (\bibinfo {year} {2016})}\BibitemShut {NoStop}%
\bibitem [{\citenamefont {Frano}\ \emph {et~al.}(2020)\citenamefont {Frano},
  \citenamefont {Blanco-Canosa}, \citenamefont {Keimer},\ and\ \citenamefont
  {Birgeneau}}]{frano_charge_2020}%
  \BibitemOpen
  \bibfield  {author} {\bibinfo {author} {\bibfnamefont {A.}~\bibnamefont
  {Frano}}, \bibinfo {author} {\bibfnamefont {S.}~\bibnamefont
  {Blanco-Canosa}}, \bibinfo {author} {\bibfnamefont {B.}~\bibnamefont
  {Keimer}},\ and\ \bibinfo {author} {\bibfnamefont {R.~J.}\ \bibnamefont
  {Birgeneau}},\ }\href {https://doi.org/10.1088/1361-648X/ab6140} {\bibfield
  {journal} {\bibinfo  {journal} {Journal of Physics: Condensed Matter}\
  }\textbf {\bibinfo {volume} {32}},\ \bibinfo {pages} {374005} (\bibinfo
  {year} {2020})},\ \bibinfo {note} {publisher: IOP Publishing}\BibitemShut
  {NoStop}%
\bibitem [{\citenamefont {Rossi}\ \emph {et~al.}(2023)\citenamefont {Rossi},
  \citenamefont {Johnson}, \citenamefont {Balgley}, \citenamefont {Thomas},
  \citenamefont {Francaviglia}, \citenamefont {Dettori}, \citenamefont
  {Schmid}, \citenamefont {Watanabe}, \citenamefont {Taniguchi}, \citenamefont
  {Cothrine}, \citenamefont {Mandrus}, \citenamefont {Jozwiak}, \citenamefont
  {Bostwick}, \citenamefont {Henriksen}, \citenamefont {Weber-Bargioni},\ and\
  \citenamefont {Rotenberg}}]{rossi_direct_2023}%
  \BibitemOpen
  \bibfield  {author} {\bibinfo {author} {\bibfnamefont {A.}~\bibnamefont
  {Rossi}}, \bibinfo {author} {\bibfnamefont {C.}~\bibnamefont {Johnson}},
  \bibinfo {author} {\bibfnamefont {J.}~\bibnamefont {Balgley}}, \bibinfo
  {author} {\bibfnamefont {J.~C.}\ \bibnamefont {Thomas}}, \bibinfo {author}
  {\bibfnamefont {L.}~\bibnamefont {Francaviglia}}, \bibinfo {author}
  {\bibfnamefont {R.}~\bibnamefont {Dettori}}, \bibinfo {author} {\bibfnamefont
  {A.~K.}\ \bibnamefont {Schmid}}, \bibinfo {author} {\bibfnamefont
  {K.}~\bibnamefont {Watanabe}}, \bibinfo {author} {\bibfnamefont
  {T.}~\bibnamefont {Taniguchi}}, \bibinfo {author} {\bibfnamefont
  {M.}~\bibnamefont {Cothrine}}, \bibinfo {author} {\bibfnamefont {D.~G.}\
  \bibnamefont {Mandrus}}, \bibinfo {author} {\bibfnamefont {C.}~\bibnamefont
  {Jozwiak}}, \bibinfo {author} {\bibfnamefont {A.}~\bibnamefont {Bostwick}},
  \bibinfo {author} {\bibfnamefont {E.~A.}\ \bibnamefont {Henriksen}}, \bibinfo
  {author} {\bibfnamefont {A.}~\bibnamefont {Weber-Bargioni}},\ and\ \bibinfo
  {author} {\bibfnamefont {E.}~\bibnamefont {Rotenberg}},\ }\href
  {https://doi.org/10.1021/acs.nanolett.3c01974} {\bibfield  {journal}
  {\bibinfo  {journal} {Nano Letters}\ }\textbf {\bibinfo {volume} {23}},\
  \bibinfo {pages} {8000} (\bibinfo {year} {2023})},\ \bibinfo {note}
  {publisher: American Chemical Society}\BibitemShut {NoStop}%
\bibitem [{\citenamefont {Liu}\ \emph {et~al.}(2023)\citenamefont {Liu},
  \citenamefont {Yang}, \citenamefont {Wang}, \citenamefont {Lu}, \citenamefont
  {Ma},\ and\ \citenamefont {Wu}}]{Liu_PhysRevB.107.165134}%
  \BibitemOpen
  \bibfield  {author} {\bibinfo {author} {\bibfnamefont {L.}~\bibnamefont
  {Liu}}, \bibinfo {author} {\bibfnamefont {K.}~\bibnamefont {Yang}}, \bibinfo
  {author} {\bibfnamefont {G.}~\bibnamefont {Wang}}, \bibinfo {author}
  {\bibfnamefont {D.}~\bibnamefont {Lu}}, \bibinfo {author} {\bibfnamefont
  {Y.}~\bibnamefont {Ma}},\ and\ \bibinfo {author} {\bibfnamefont
  {H.}~\bibnamefont {Wu}},\ }\href
  {https://doi.org/10.1103/PhysRevB.107.165134} {\bibfield  {journal} {\bibinfo
   {journal} {Phys. Rev. B}\ }\textbf {\bibinfo {volume} {107}},\ \bibinfo
  {pages} {165134} (\bibinfo {year} {2023})}\BibitemShut {NoStop}%
\end{thebibliography}%
\end{document}